\documentclass[11pt]{article}
\pdfoutput=1

\usepackage{a4wide}
\usepackage[]{amsmath}

\usepackage{fourier} 
\usepackage[scaled=0.875]{helvet} 

\usepackage{tikz}
\usetikzlibrary{matrix}

\tikzset{
  pretableaumatrix/.style={
    ampersand replacement=\&,
    matrix of math nodes,
    outer sep=1mm,
    inner sep=0mm,
    anchor=center,
    row sep={between borders,-\pgflinewidth},
    column sep={between borders,-\pgflinewidth},
    dottedentry/.style={densely dotted},
    dashedentry/.style={densely dashed},
    spaceentry/.style={draw=none,execute at begin node=\null},
  },
  pretableaunode/.style={
    font=\small,
    draw=gray,
    sharp corners,
    rectangle,
    anchor=base,
    text height=3.75mm,
    text depth=1.25mm,
    minimum height=5mm,
    minimum width=5mm,
    inner sep=0mm,
    outer sep=0mm,
    doublewidth/.style={minimum width=10mm},
    footnotesize/.style={font=\footnotesize},
    scriptsize/.style={font=\scriptsize},
  },
  tableaumatrix/.style={
    pretableaumatrix,
    every node/.append style={
      pretableaunode,
    },
  },
  medtableaumatrix/.style={
    pretableaumatrix,
    every node/.append style={
      pretableaunode,
      font=\footnotesize,
      text height=2.75mm,
      text depth=.75mm,
      minimum height=3.5mm,
      minimum width=3.5mm
    },
  },
  smalltableaumatrix/.style={
    pretableaumatrix,
    every node/.append style={
      pretableaunode,
      font=\scriptsize,
      text height=1.85mm,
      text depth=.15mm,
      minimum height=2.5mm,
      minimum width=2.5mm,
    },
  },
  tinytableaumatrix/.style={
    pretableaumatrix,
    every node/.append style={
      pretableaunode,
      font=\tiny,
      text height=1.25mm,
      text depth=.15mm,
      minimum height=1.75mm,
      minimum width=1.75mm
    },
  },
  tableau/.style={
    baseline=-1.25mm,
    every matrix/.style={tableaumatrix},
  },
  medtableau/.style={
    baseline=-1.25mm,
    every matrix/.style={medtableaumatrix},
  },
  smalltableau/.style={
    baseline=-1.25mm,
    every matrix/.style={smalltableaumatrix},
  },
  preshapetableaumatrix/.style={
    pretableaumatrix,
    execute at end cell={\strut},
    every node/.append style={
      draw=black,
      anchor=base,
      inner sep=0mm,
      outer sep=0mm,
    },
    shadedentry/.style={fill=gray},
    darkshadedentry/.style={fill=darkgray},
  },
  medshapetableaumatrix/.style={
    preshapetableaumatrix,
    every node/.append style={
      text height=2.75mm,
      text depth=.75mm,
      minimum height=3.5mm,
      minimum width=3.5mm
    },
  },
  shapetableaumatrix/.style={
    ampersand replacement=\&,
    matrix of math nodes,
    outer sep=0mm,
    inner sep=0mm,
    anchor=base,
    row sep={between borders,-\pgflinewidth},
    column sep={between borders,-\pgflinewidth},
    execute at begin cell={\strut},
    every node/.append style={draw,anchor=base,text height=1mm,text depth=.5mm,minimum size=1.5mm,inner sep=0mm,outer sep=0mm},
  },
  shapetableau/.style={
    every matrix/.style={shapetableaumatrix},
  },
  topalign/.style={
    every matrix/.append style={name=maintableau,anchor=maintableau-1-1.base},
    baseline,
  },
}

\newcommand*\smalltableau[2][]{\tikz[smalltableau,#1]\matrix{#2};}
\usepackage{amsmath,amssymb,amsxtra,amsthm}

\usepackage{mathrsfs}

\usepackage{setspace}
\usepackage{array}
\usepackage{booktabs}

\usepackage{colortbl}
\usepackage{xcolor}
\colorlet{titlerowcolor}{gray!15}

\usepackage[
        backend=bibtex,
        sorting=none,
        style=phys,
        eprint=true,
        doi=false,
        biblabel=brackets
        ]{biblatex}

\bibliography{KBbackground}

\definecolor{blue3}{RGB}{31,119,180}
\definecolor{red3}{RGB}{214,39,40}
\definecolor{orange3}{RGB}{255,127,14}
\definecolor{green3}{RGB}{44,160,44}

\usepackage{colortbl}
\definecolor{lightgreen}{cmyk}{0.2, 0, 0.2, 0.2}
\definecolor{lightgray}{cmyk}{0.1,0.2,0,0.1}
\definecolor{lightgray2}{cmyk}{0.1,0.1,0,0.1}

\usepackage{hyperref}
\hypersetup{colorlinks=true,linkcolor=red3,citecolor=green3,urlcolor=teal,pdfencoding=auto}

\numberwithin{equation}{section}
\numberwithin{table}{section}
\numberwithin{figure}{section}

\author{
  \begin{minipage}{1.00\linewidth}
    \vspace{1cm}
    \begin{center}
      \begin{small}
        \textbf{ Giorgio Leone}
        \\ \vspace{1cm}
        {\em Scuola Normale Superiore and INFN}
        \\
        {\em Piazza dei Cavalieri 7, 56126, Pisa, Italy}
     \end{small}
    \end{center}
    \vspace{1cm}
  \end{minipage}
}

\date{}

\title{\vspace{3cm}
  \begin{huge} \textbf{New comments on six-dimensional orientifold vacua with reduced rank and unitarity constraints} 	
  \end{huge}
  \\ \vspace{.7cm}
}

\begin{document}

\begin{titlepage}
  \maketitle
  \thispagestyle{empty}

  \vspace{-14cm}
  \begin{flushright}
   \end{flushright}

  \vspace{11cm}

  \begin{center}
    \textsc{Abstract}\\
  \end{center}

\noindent We revisit and extend the construction of six-dimensional orientifolds built upon the $T^4/\mathbb{Z}_N$ orbifolds with a non-vanishing Kalb-Ramond background, both in the presence of $\mathcal{N}=(1,0)$ supersymmetry and Brane Supersymmetry Breaking, thus amending some statements present in the literature. In the $N=2$ case, we show how the gauge groups on unoriented D9 and D5 (anti-)branes do not need to be correlated, but can be independently chosen complex or real. For $N>2$ we find that the Diophantine tadpole conditions severely constrain the vacua. Indeed, only the $N=4$ orbifold with a rank-two Kalb-Ramond background may admit integer solutions for the Chan-Paton multiplicities, if the $\mathbb{Z}_4$ fixed points support $\text{O}5_-$ planes, both with and without supersymmetry. All other cases would involve a fractional number of branes, which is clearly unacceptable. We check the consistency of the new $\mathbb{Z}_2$ and $\mathbb{Z}_4$ vacua by verifying the unitarity constraints for string defects coupled to Ramond-Ramond two-forms entering the Green-Schwarz-Sagnotti mechanism.

\vfill

{\small
\begin{itemize}
\item[E-mail:]  {\tt giorgio.leone@sns.it}

\end{itemize}
}

\end{titlepage}

\setstretch{1.1}


{		\hypersetup{linkcolor=black}
		\tableofcontents	}

\newpage

\section{Introduction}

The string landscape is a wild land full of surprises.  Indeed, the corners admitting an exact world-sheet description are very few and, in spite of a vast literature, even those hide interesting features yet to be uncovered. In the present note, we revisit the construction of the well-studied K3 orientifolds, with and without space-time supersymmetry, with a discrete Kalb-Ramond background, intending to complete the study present in the literature \cite{Angelantonj:1999jh, Kakushadze:1998bw, Kakushadze:2000hm} and, hopefully, provide a complete characterisation of such vacua.

\noindent These backgrounds admit an underlying conformal field theory (CFT) which is exactly solvable and affords a complete string theory analysis. The pioneering work \cite{Bianchi:1991eu} showed, for instance, that although $B$-field is odd under the world-sheet parity $\Omega$, a discrete background is actually allowed. Such a background reduces the Ramond-Ramond (R-R) charge of the O9 planes which then require fewer D9 branes to cancel the tadpoles, resulting in a gauge group of reduced rank. Moreover, Wilson lines may interpolate between orthogonal and symplectic gauge groups, passing through unitary ones  \cite{Bianchi:1991eu}. Upon T-dualities, the reduction of the charge of O9 planes admits a suggestive geometric interpretation in terms of the simultaneous presence of $\text{O}_+$ and $\text{O}_-$ planes of reduced dimension \cite{Witten:1997bs}, while the $B$-field can be interpreted as a characteristic class in a $\mathbb{Z}_2$ cohomology obstructing the presence of bundles admitting a vector structure \cite{Witten:1997bs, Bianchi:1997rf, Bachas:2008jv} (see \cite{Sen:1997pm} for a world-sheet argument).

\noindent For orientifolds based on toroidal orbifolds \cite{Dixon:1985jw, Dixon:1986jc} a rank $b$ background accommodates a richer scenario. Not only both $\text{O}_-$  and $\text{O}_+$ planes are simultaneously present, thus requiring a smaller number of D9 and D5 branes to cancel tadpoles, but open strings with Neumann-Dirichlet boundary conditions come in multiple families \cite{Bianchi:1990yu, Angelantonj:1996mw, Angelantonj:1999jh, Kakushadze:1998bw}. When combined with the Brane Supersymmetry Breaking (BSB) involution, which flips the charge and tension of O5 planes sitting on the $\mathbb{Z}_2$ orbifold fixed points, many configurations involving a varying number of O$_+$ and O$_-$ planes are actually possible \cite{Angelantonj:1999jh}.

\noindent In \cite{Angelantonj:1999jh}, it was argued that only the $T^4/\mathbb{Z}_2$ orientifolds were compatible with a Kalb-Ramond background, and D9 and D5 branes would both support unitary groups or orthogonal (for D9) and symplectic (for D5) ones, these two configurations being connected by Wilson lines and brane displacements. On the other hand, \cite{Kakushadze:1998bw} argued that a Kalb-Ramond background is compatible with any K3 orientifold and few vacua were guessed most of which however turned out to be anomalous, since R-R tadpoles were not properly cancelled. 

\noindent Recently, new solutions of tadpole conditions for the $\mathbb{Z}_2$ orientifolds with BSB were found, which resulted in brane configurations with mixed unitary and real gauge groups \cite{Angelantonj:2024iwi}, which was in principle forbidden by the arguments of \cite{Angelantonj:1999jh, Kakushadze:1998bw}. The possibility of extending the (incomplete) study of \cite{Angelantonj:1999jh, Kakushadze:1998bw} is the main motivation of this work. 

\noindent We start by revisiting the simple $T^4 /\mathbb{Z}_2$ supersymmetric orientifold with a background field of rank $b=2,4$. In agreement with the results of \cite{Angelantonj:2024iwi}, D9 and D5 branes can independently carry real or complex Chan-Paton charges resulting in various combinations of orthogonal, symplectic and unitary gauge groups. In the case when unitary groups are supported on the D9 branes while real groups are associated to D5 branes, or viceversa, D9-D5 strings do not feel the orbifold projection, as indeed required by a consistent tadpole cancellation (in the transverse channel amplitudes) and particle interpretation (in the direct channel amplitudes).

\noindent We then turn to the other $\mathbb{Z}_N$ orientifolds with $N=3,4,6$. Actually, in \cite{Angelantonj:1999jh} it was already shown that the $N=3$ case does not allow for such a discrete deformation, while \cite{Kakushadze:1998bw} suggested that they could be allowed in the $N=4$ and $6$ cases. However, it did not provide a detailed realisation of the vacua, and indeed most of the solutions presented there turned out to be inconsistent. These orbifolds are quite complicated to study because they involve fixed points of different type, which not necessarily coincide with the location of O5 planes, which are associated to the fixed points of $\Omega g^{N/2}$, $g$ being the generator of $\mathbb{Z}_N$. In fact, if the parametrisation of the Chan-Paton charges are chosen so that the direct-channel annulus and M\"obius strip amplitudes provide a consistent particle interpretation, untwisted and twisted R-R tadpoles are uniquely determined, but not necessarily admit acceptable solutions. This is indeed what happens for the twisted tadpoles for the case $N=6$ and rank $b=2,4$ and for the case $N=4$ with rank $b=4$, both for supersymmetric and BSB constructions. Indeed, in these set-ups a correct interpretation of the annulus amplitude in the transverse channel requires a fractional number of D5 or $\overline{\text{D5}}$ branes, leading to unavoidable inconsistencies.
The only situation where a meaningful solution can be found is $\mathbb{Z}_4$ with a rank-two $B$-field background. These constructions involve twelve (four) $O5_-$ and four (twelve) $O5_+$ planes in the supersymmetric (BSB) vacua which, in principle, could be arbitrarily distributed among the four $\mathbb{Z}_4$ fixed points. However, it can be shown that only when each $\mathbb{Z}_4$ fixed point supports an $O5_-$ plane a solution to the twisted tadpole conditions can be found which is compatible with a consistent particle interpretation of the spectrum, both in the supersymmetric and BSB cases. In fact, in the other cases, O5$_+$ planes on the $\mathbb{Z}_4$ fixed points carry a non-trivial twisted R-R charge which must be cancelled locally and would require D5 branes with non-integer R-R charge, which is clearly inconsistent.

\noindent Although these constructions may sound exotic, they correspond to {\em bona fide} vacua since local anomalies are cancelled by a generalised Green-Schwarz-Sagnotti mechanism \cite{Green:1984bx, Green:1984sg, Sagnotti:1992qw} and the anomaly inflow on one-dimensional defects is cancelled by a unitary CFT living on their world-sheet \cite{Kim:2019vuc, Angelantonj:2020pyr, Angelantonj:2024iwi}.

\noindent The interest on these orientifold vacua with a non-vanishing Kalb-Ramond background is not purely academic. In fact, it was shown in \cite{Angelantonj:2000qtl} that a non-trivial $B$-field background on $T^4 /\mathbb{Z}_2$ orientifolds with magnetised branes is instrumental to obtain an odd number of families of chiral fermions. Extending this analysis to the $\mathbb{Z}_N$ orientifolds with magnetic fields with or without Kalb-Ramond background is still an interesting open problem with potential phenomenological interest, to which we hope to return in the near future. In fact, although T-duality maps magnetised branes on a torus to intersecting ones, and a rank $b$ $B$-field to a (discrete) non-diagonal metric for the compactification torus \cite{Angelantonj:1999xf}, on orbifolds things are more complicated. Aside from the $\mathbb{Z}_2$ case, upon T-duality, a geometric action, where left and right movers are rotated by the same angle, turns into a non-geometric one, where left and right moving coordinates undergo different rotations. As a result, magnetic fields and non-trivial angles on geometric orbifolds describe completely different vacua.

\noindent The paper is organised as follows. In Section \ref{Sec:6dorientifold}, a general description of the orientifold amplitudes to be used in the rest of the paper is provided. In Section \ref{SSec:Z2B} the additional solutions with a non-trivial $B$-field for the $T^4/\mathbb{Z}_2$ orbifold are discussed by straightforwardly extending the observations in \cite{Angelantonj:2024iwi} to the supersymmetric case. In Section \ref{SSec:Z4BSUSY} the only available discrete deformation of the $T^4/\mathbb{Z}_4$ supersymmetric orientifold with a rank $2$ Kalb-Ramond field is described, while in Section \ref{SSec:Z4BBSB} the same analysis is repeated for the BSB case. In Section \ref{SSec:discussion} the obstructions for the realisation of different scenarios are presented, while the consistency checks from the unitarity conditions on string probes are discussed in Section \ref{Sec:defects} for two examples. Finally, Section \ref{Sec:Conclusions} provides our conclusions and  outlooks. Appendix \ref{App:ZN} collects useful definitions and formul{\ae} used throughout the paper.

\section{Six dimensional orientifold vacua with a non vanishing Kalb-Ramond background field} \label{Sec:6dorientifold}

The possibility of introducing a Kalb-Ramond field for orientifold vacua \cite{Sagnotti:1987tw, Pradisi:1988xd, Horava:1989vt, Bianchi:1990yu, Bianchi:1990tb, Bianchi:1991eu, Bianchi:1997rf, Angelantonj:1999xf, Dudas:2000bn, Angelantonj:2002ct} has been intensively discussed in the literature \cite{Bianchi:1991eu, Witten:1997bs, Sen:1997pm, Kakushadze:1998bw, Angelantonj:1999jh, Kakushadze:2000hm, Bachas:2008jv}. The $B$-field is odd under the action of the orientifold projection $\varOmega$, implying that the quantum fluctuations be projected away, but still allowing background values to be present in toroidal compactifications \cite{Bianchi:1991eu}. In this setting, the momenta lying in the Narain lattice for each $T^2$ torus read \cite{Narain:1985jj, Narain:1986am}
\begin{equation}
    \begin{aligned} \label{eq:leftrightmomenta}
        & p_L^a= g^{ab} m_b + \tfrac{1}{\alpha'}( g_{ab} -  B_{ab})n^b \, ,
        \\
        & p_R^a= g^{ab} m_b - \tfrac{1}{\alpha'}( g_{ab} +  B_{ab})n^b \, ,
    \end{aligned}
\end{equation}
and the condition for world-sheet parity symmetry requires $\frac{2}{\alpha'} B_{ab} n^b$ to be an integer, thus forcing the background value for the Kalb-Ramond field to be semi-integer valued. All in all, the parent closed string partition function reads
\begin{equation}
{\mathcal T}_\text{IIB} = \tfrac{1}{N} \sum_{\alpha, \beta =0}^{N-1} n_{\alpha , \beta}\, T_B\big[ {\textstyle{\alpha \atop \beta}}\big] \, \bar T_B \big[{\textstyle{-\alpha \atop -\beta}}\big]\,\varLambda \big[{\textstyle{\alpha \atop \beta}}\big]\,,
\label{torusZN}
\end{equation}
with
\begin{equation}
\varLambda \big[{\textstyle{\alpha \atop \beta}}\big] = \begin{cases} \varLambda_{(4,4)} & \text{for} \quad \alpha = \beta = 0\,,
\\
1 & \text{otherwise}\,,
\end{cases}
\end{equation}
where $\varLambda_{(4,4)}$ denotes the Narain lattice with respect to the momenta and winding defined in \eqref{eq:leftrightmomenta}.
Furthermore, in the definition of \eqref{torusZN}, we have employed the modular blocks $T_B\big[ {\textstyle{\alpha \atop \beta}}\big]$ described in detail in Appendix \ref{App:ZN} and the multiplicity of the twisted sectors $n_{\alpha, \beta}$ corresponding to the number of fixed points under the orbifold action given by the Lefschetz theorem \cite{Lefschetz:1926, Lefschetz:1937} when $\alpha \neq 0$, while $n_{0,\beta}=1$.

\noindent Starting from \eqref{torusZN}, gauging the world-sheet parity symmetry leads to the Klein bottle amplitude encoding the action on the closed-string states of the orientifold projection \cite{Sagnotti:1987tw}. For $(\alpha, \beta)=(0,0)$, the Klein bottle amplitude encodes the contribution arising from the toroidal compactification and hence it is unaffected by the presence of the Kalb-Ramond field since the orientifold projection identifies only the states satisfying $p_L=p_R$. However, for even $N$, the orbifold point group admits the presence of an order $2$ element $g^{N/2}$ which selects states satisfying $p_L=-p_R$, and thus requires
\begin{equation} \label{eq:windingcond}
    \frac{2}{\alpha'} B_{ab} n^b \in 2 \mathbb{Z} \, .
\end{equation}
Furthermore, on the twisted two-cycles the rank $b$ Kalb-Ramond field induces an additional involution $(-1)^\sigma$ flipping the sign of the orientifold projection acting on states in the $g^{N/2}$-twisted sector. Even cycles imply the presence of $\mathcal{N}=(1,0)$ hyper multiplets in $6d$ in the massless spectrum, while odd cycles lead to  $\mathcal{N}=(1,0)$ tensor multiplets in $6d$. Such a different behaviour reflects the presence of O5$_-$ planes placed on even $\mathbb{Z}_2$-fixed points, while the odd ones support O5$_+$ planes. 

\noindent The multiplicities appearing in the twisted sector projected by $g^{2\beta}$ are given by the sum of the number of fixed points weighted by the action of the involution and will be denoted in the following as $\Delta_{N/2, \beta}$, to be determined by imposing a consistent interpretation of the one-loop and tree-level amplitudes connected by suitable modular transformations. 

\noindent As a result, the Klein bottle amplitude in the direct channel reads
\begin{equation}
{\mathcal{K}}^{(b)}= \tfrac{1}{N}  \sum_{\beta=0}^{N-1} T_B \big[ {\textstyle{0 \atop 2 \beta }}\big] \, \varGamma_\beta^{(b)} + \, \tfrac{1}{N}\sum_{\beta=0}^{N-1} \Delta_{N/2, \beta}\, T_B \big[ {\textstyle{N/2 \atop 2 \beta}}\big]\,,
\label{eq:KleinZNB}
\end{equation}
where the lattice terms are given by 
\begin{equation} \label{eq:latticeterm}
    \varGamma_\beta^{(b)}=\begin{cases}
        \displaystyle{\sum_{\{m \} }} q^{\frac{\alpha'}{2} m^\intercal \, g^{-1} \, m  } \equiv P(0,0) \, , & \qquad \beta=0 \, ,
        \\
       \displaystyle{ \sum_{\{n \} }}  \displaystyle{\sum_{\{ \epsilon \in \{ 0,1 \} \} }}  q^{\frac{1}{2\alpha'} n^\intercal \, g \, n} \,  e^{\frac{2 \pi i}{\alpha'} n^\intercal B \epsilon} \equiv W(0,B) \, , & \qquad \beta=N/2 \, ,
        \\
        1 \, , & \qquad \text{otherwise} \, .
    \end{cases} 
\end{equation}
In the expressions above, we have adopted the convention for which the first argument denotes the shift of the momenta or winding, while the second one encodes the presence of a non-trivial projection. The tree-level propagation is straightforwardly obtained by performing an $S$ modular transformation on \eqref{eq:KleinZNB}, giving
\begin{equation} \label{eq:kleinZNBtr}
\begin{aligned}
    \tilde{\mathcal{K}}^{(b)}= & \frac{2^5}{N} \left \{ T_B \big[ {\textstyle{0 \atop 0 }}\big] \left ( \widetilde{W}(0,0) + \widetilde{P}(B,0) \right ) +  \, \frac{\Delta_{N/2,0}+\Delta_{N/2,N/2}}{16} T_B \big[ {\textstyle{0 \atop N/2 }}\big] \right \} 
    \\
    &+ \frac{2^4}{N} \sum_{\beta=0}^{N-1}\left \{ T_B \big[ {\textstyle{2 \beta \atop 0 }}\big] \, \left ( 2 \sin 2 \beta \pi /N \right )^2 -  \, T_B \big[ {\textstyle{2 \beta \atop N/2 }}\big] \frac{\Delta_{N/2,\beta}+\Delta_{N/2,N-\beta}}{2} \right \} \, ,
\end{aligned}
\end{equation}
where $ \widetilde{W}(0,0)$ and $  \widetilde{P}(B,0)$ are obtained via the Poisson resummation of $P(0,0)$ and $W(0,B)$ defined in \eqref{eq:latticeterm}, respectively.

\noindent Enforcing a consistent interpretation in the tree-level channel of closed string states propagating between pairs of orientifold planes \cite{Bianchi:1990yu} imposes a first condition on the definition of the degeneracies $\Delta_{N/2,\beta}$. Indeed, we shall require the massless contribution to factorise into a sum of squares \cite{Angelantonj:1999jh} in the tree-level channel, so to enforce a meaningful interpretation of orientifold planes as physical objects. 

\noindent Starting from the untwisted sector we shall impose  
\begin{equation}
\Delta_{N/2,0}=\Delta_{N/2,N/2}= \zeta \cdot 2^{4-b/2} \, ,
\end{equation}
where
\begin{equation}
    \zeta= \text{sign} \left ( \sum_{p \, \in \, \, \text{f.p.} \left \{\mathbb{Z}_2 \right \}} (-1)^{\sigma_p} \right ) \, ,
\end{equation}
and $\sigma_p$ is the eigenvalue of the associated $\mathbb{Z}_2$ fixed point under $\sigma$. As a result, using the definitions reported in Appendix \ref{App:ZN}, we see that the massless contribution to the untwisted piece encoding the tension and charge of the orientifold planes reads
\begin{equation} \label{eq:kleinZNBtrunmassless}
    \tilde{\mathcal{K}}^{(b)}_{\text{untw}} \sim \frac{2^{-5}}{2} \left \{ \tau^B_{0,0} \left (- 2^5 \sqrt{v} - \zeta \, \frac{2^{5-b/2}}{\sqrt{v}} \right )^2 + \ldots \right \} \, .
\end{equation}

\noindent Two comments are in order: the presence of rank $b$ Kalb-Ramond field implies a smaller overall tension and charge of orientifold planes reflected in the simultaneous presence of O5$_+$ and O5$_-$ planes, which indeed behave differently under the orientifold projection. Furthermore, the $\zeta$ is ultimately responsible for the nature of D-branes required to cancel anomalies. Indeed, whenever $\zeta=1$ we have to introduce D5 branes leading to supersymmetric spectra, while if $\zeta=-1$ supersymmetry is broken {\em à la} BSB since $\overline{\text{D5}}$ branes have now to be introduced. 

\noindent The propagation of $2 \beta$-twisted closed string states between orientifold planes is encoded into 
\begin{equation}\label{eq:kleinZNBtrtw}
    \tilde{\mathcal{K}}^{(b)}_{2\beta-\text{tw}}= \frac{2^4}{N} \left \{ T_B \big[ {\textstyle{2 \beta \atop 0 }}\big] \, \left ( 2 \sin 2 \beta \pi /N \right )^2 -  \, T_B \big[ {\textstyle{2 \beta \atop N/2 }}\big] \frac{\Delta_{N/2,\beta}+\Delta_{N/2,N-\beta}}{2} \right \} \, ,
\end{equation}
where the $2\beta$-twisted charge of orientifold planes thus depends on the values of the degeneracies $\Delta_{N/2,\beta}$. The consistency of the supersymmetric and BSB constructions with $b=0$ actually requires
\begin{equation} \label{eq:degkleinZN}
    \Delta_{N/2,\beta}= \zeta \, (2 \sin \pi \beta/N)^4 \, ,
\end{equation}
but in the presence of a non-vanishing $B$-field, more values are {\em a priori} allowed, compatibly with a consistent particle interpretation in the closed and open-string sectors. Looking at the partition function \eqref{eq:KleinZNB} and defining $n_\mp$($N_\mp$) the number of the $\mathbb{Z}_N$ fixed points ($\mathbb{Z}_2$ fixed points organised into $N/2$-uplets) supporting O5$_\mp$ planes, we shall require
\begin{equation} \label{eq:ZNfixedpoints}
    n_{\mp}= \frac{(2 \sin \pi /N)^4 \pm \Delta_{N/2,1}}{2}  \in \, \mathbb{N} \, ,
\end{equation}
and
\begin{equation} \label{eq:Z2fixedpoints}
   N_{\mp}= \frac{ 16 (1 \pm \zeta \cdot 2^{-b/2} ) - \Delta_{2,1} (1 \pm 1) }{N}  \ \  \in \, \mathbb{N}  \, .
\end{equation}
The conditions \eqref{eq:ZNfixedpoints} and \eqref{eq:Z2fixedpoints} constrain the possible involutions $(-1)^\sigma$ at our disposal and correspond to the allowed values to enforce physically meaningful multiplicities. These considerations are compatible with the case of a trivial Kalb-Ramond field $b=0$. If $\zeta=1$ the $g^{N/2}$-twisted states are all symmetrised, thus recovering the spectrum of supersymmetric vacua, while if $\zeta=-1$ the $g^{N/2}$-twisted states are all antisymmetrised, thus reproducing the spectrum of BSB constructions. 

\noindent To cancel irreducible anomalies \cite{Aldazabal:1999nu, Bianchi:2000de}, D9 and D5 or $\overline{\text{D5}}$ branes\footnote{Actually, the construction can accommodate $\overline{\text{D9}}$ as well, realising the so-called Sugimoto vacua which unavoidably break supersymmetry. Although straightforward, we will ignore this further possibility in the rest of the paper.} have to be introduced depending on the value of $\zeta$. Concerning D9/D9 oriented amplitudes, the $B$-field does play a non-trivial role since the winding states propagating in the transverse channel satisfy $p_L=-p_R$ and hence enforce \eqref{eq:windingcond}. As a result, indicating as $N_\beta$ ($D_{(k),\beta}$) the Chan-Paton labels projected by $\beta$ for D9 (D5 or $\overline{\text{D5}}$) branes lying on the $k$-th fixed point, $y_k$, the annulus partition function in the transverse channel reads
\begin{equation} \label{eq:transverseannulusZN}
    \begin{aligned}
        \tilde{\mathcal{A}}^{(b)}_\zeta & =  \frac{2^{-5}}{N}  \Bigg [ \bigg ( 2^b \, N_0^2 \, \widetilde{W}(0,B)  + \sum_{k, \ell} D_{(k),0} D_{(\ell),0} \, \widetilde{P}(0,y_k-y_\ell) \bigg )   T_B \big[ {\textstyle{0 \atop 0}}\big] 
        \\
        &\qquad \qquad + 2 \cdot 2^{b/2} \, N_0 \sum_k D_{(k),0} \,  \left ( B_B \big[ {\textstyle{0 \atop N/2}}\big] - \zeta F_B \big[ {\textstyle{0 \atop N/2}}\big] \right ) \Bigg ]
        \\
        & + \frac{2^{-3}}{N} \sum_{\beta=1}^{N-1} \Bigg [ \big (2 \sin \pi \beta/N \big )^2  \bigg (  N_\beta^2 \,  + \sum_k D_{(k),\beta}^2  \bigg ) T_B \big[ {\textstyle{\beta \atop 0}}\big] 
        \\
        & \qquad  \qquad  - 2 \cdot 2^{b/2} \, N_\beta \sum_k D_{(k),\beta}  \, \bigg ( B_B \big[ {\textstyle{\beta \atop N/2}}\big] - \zeta F_B \big[ {\textstyle{\beta \atop N/2}}\big] \bigg )  \Bigg ] \, ,
    \end{aligned}
\end{equation}
where we have separated the contributions of space-time bosons and fermions in $B_B\big[ {\textstyle{a\atop b}}\big]$ and $F_B\big[ {\textstyle{a \atop b}}\big]$, reported in Appendix \ref{App:ZN}.

\noindent Notice that the sign $\zeta$ determines the R-R charge since it appears in front of the R-R sector $F_B \big[ {\textstyle{\beta \atop N/2}}\big]$ and one can read from the massless states the relative sign between D9 and five-branes charges, corresponding to D5 branes or $\overline{\text{D5}}$ branes if $\zeta=1$ or $\zeta=-1$.

\noindent Finally, we can describe the propagation of closed string states between boundaries and cross-caps encoded in the transverse channel of the M\"obius strip amplitude, which yields
\begin{equation} \label{eq:transversemoebiusZN}
    \begin{aligned}
        \tilde{\mathcal{M}}^{(b)}_\zeta = -\frac{2}{N} & \Bigg \{  N_0 \bigg (  T_B \big[ {\textstyle{0 \atop 0}}\big]  \, 2^{\frac{4-b}{2}} \, \widetilde{W}^{(\gamma)}(0,B) + \zeta \,   T_B \big[ {\textstyle{0 \atop N/2}}\big]   \bigg ) 
        \\
        & + \sum_{k} D_{(k),0}  \bigg ( \, 2^{\frac42} \, \widetilde{P}^{(\tilde \gamma)}(B,0) \left( \zeta B_B\big[ {\textstyle{0 \atop 0}}\big] -  F_B \big[ {\textstyle{0 \atop 0}}\big] \right) + \zeta \, \left( \zeta B_B\big[ {\textstyle{0 \atop N/2}}\big] -  F_B \big[ {\textstyle{0 \atop N/2}}\big] \right) \bigg ) 
        \\
        &  + \xi \sum_{\beta=1}^{N/2-1}  N_{2 \beta} \bigg [ \big (2 \sin  \pi \beta/N \big )^2 T_B \big[ {\textstyle{2 \beta \atop \beta }}\big] - \zeta \big (2 \sin  \pi (\beta+N/2)/N \big )^2 T_B \big[ {\textstyle{2 \beta \atop \beta + N/2 }}\big] \bigg ]
        \\
        &+ {\tilde \xi} \sum_{\beta=1}^{N/2-1} \sum_k D_{(k), 2 \beta} \bigg [ \big (2 \sin  \pi \beta/N \big )^2 \left( \zeta B_B\big[ {\textstyle{2 \beta \atop \beta }}\big] -  F_B \big[ {\textstyle{2 \beta \atop \beta }}\big] \right)  \\
        & \qquad \qquad \qquad \qquad - \zeta \big (2 \sin  \pi (\beta+N/2)/N \big )^2 \left( \zeta B_B\big[ {\textstyle{2 \beta \atop \beta + N/2}}\big] -  F_B \big[ {\textstyle{2 \beta \atop \beta + N/2}}\big] \right)  \, \bigg ] \Bigg \} \, ,
    \end{aligned}
\end{equation}
where we have defined the lattice term contributions
\begin{equation}
\begin{split}
   & \widetilde{W}^{(\gamma)}(0,B) = 2^{-2} \sum_{\{n \} }  \sum_{\{ \epsilon \in \{ 0,1 \} \} }  \gamma_{\epsilon}\, q^{\frac{1}{4 \alpha'} n^\intercal \, g \, n} \,  e^{\frac{2 \pi i}{\alpha'} n^\intercal B \zeta} \, ,
   \\
   & \widetilde{P}^{(\tilde{\gamma})}(B,0) =  2^{-2} \sum_{\{ m \} }  \sum_{\{ \epsilon \in \{ 0,1 \} \} }  \tilde \gamma_{\epsilon} \, q^{\frac{\alpha'}{4} (m+ \frac{1}{\alpha'} B \zeta )^\intercal \, g^{-1} \, (m+ \frac{1}{\alpha'} B \zeta ) }\, ,
\end{split}
\end{equation}
with the signs $\xi$ and $\tilde \xi$ required to enforce a Chan-Paton decomposition to be compatible with the structure of the gauge group. In addition, the compatibility between $\tilde{ \mathcal{K}}$, $\tilde{ \mathcal{A}}$ and $\tilde{ \mathcal{M}}$ requires the introduction of signs, $\gamma$ for the momenta and ${\tilde \gamma}$ for the winding states, ascribing to the reduction of the overall charge and tension. Therefore, to interpret \eqref{eq:transversemoebiusZN} as the average between the charges of cross-caps and boundaries for the massless states it is required to enforce the conditions \cite{Angelantonj:1999jh}
\begin{equation} \label{eq:gammatr}
    \sum_{\{ \epsilon=0,1 \}} \gamma_\epsilon = 2^{4/2} \, , \qquad \sum_{\{ \epsilon \, \in \,  \text{ker}(B) \}} \tilde \gamma_\epsilon = 2^{(4-b)/2} \, .
\end{equation}

\noindent Performing an $S$ modular transformation implies the direct channel of the annulus amplitude of strings stretched between D9 and D9 branes be
\begin{equation}
{\mathcal A}_{99}^{(b)} = \tfrac{1}{N} \sum_{\beta=0}^{N-1} N_\beta^2 \, T_B \big[ {\textstyle{0 \atop \beta}}\big] \, \left ( 2^{b-4} \, P(B,0) \right )^{\delta_{\beta ,0}} \,.
\label{eq:annulus99B}
\end{equation}
Open strings with DD boundary conditions propagate in the transverse channel KK momenta with $p_L=p_R$ and thus the Kalb-Ramond field leaves the D5/D5 (or $\overline{\text{D5}}/\overline{\text{D5}}$) amplitude unchanged
\begin{equation}
{\mathcal A}^{(b)}_{ 5_{i,\zeta} 5_{\ell,\zeta}} =\frac{1}{N}\sum_{i,\ell} D_{(i)\, , \,0 } \, D_{(\ell)\, , \,0 } \, T_B \big[ {\textstyle{0 \atop 0}}\big] W (y_i - y_\ell,0 ) + \frac{1}{N} \sum_{\beta=1}^{N-1} \sum_{i} D_{(i)\, , \,\beta}^2 \, T_B \big[ {\textstyle{0 \atop \beta}}\big] \, ,
\label{eq:annulus55B}
\end{equation}
with the notation $5_{\ell,\zeta}$ to denote D$5$ branes or $\overline{\text{D}}5$ branes located at the point $y_{\ell}$ for $\zeta=1$ and $\zeta=-1$, respectively. Furthermore, requiring in the transverse channel D-branes to carry tension and charge implies a non-trivial degeneracy for open strings stretched between D9 and D5 (or $\overline{\text{D5}}$) branes. This observation then leads to the amplitude
\begin{equation}
{\mathcal A}_{9 5_{i,\zeta}}^{(b)} = \frac{2^{b/2}}{N} \sum_{\beta =0}^{N-1}  N_\beta \, D_{(i)\,,\,\beta} \, T_{ B_\zeta} \big[ {\textstyle{N/2 \atop \beta}}\big] \, ,
\label{eq:annulus95B}
\end{equation}
where we have indicated the GSO projection relevant for D5 branes $B_1=B$, while $B_{-1}=\hat B$ if $\overline{\text{D5}}$ branes are introduced.

\noindent The orientifold projection on the open sector is obtained by performing a $P$-modular transformation on \eqref{eq:transversemoebiusZN} and can be written in compact form as 
\begin{equation}
\begin{split}
{\mathcal M}_9^{(b)} &= - \frac{1}{N} \sum_{\beta=0}^{N/2-1} N_{2\beta}\, \left [ T_B \big[ {\textstyle{0 \atop \beta}}\big]\, P_\beta^{(b)}- \zeta T_B \big[ {\textstyle{0 \atop \beta+N/2}}\big] \right ] \,, 
\\
{\mathcal M}_{ 5_{i,\zeta}}^{(b)} &=-\zeta \frac{1}{N} \sum_{\beta=0}^{N-1} \sum_i D_{(i)\,,\, 2\beta} \,\Bigg [ \left( B_B\big[ {\textstyle{0 \atop \beta}}\big] - \zeta  F_B \big[ {\textstyle{0 \atop \beta}}\big] \right)\, W_\beta^{(b)} 
\\
& \qquad \qquad \qquad \qquad \qquad   - \zeta \left( B_B\big[ {\textstyle{0 \atop \beta+N/2}}\big] - \zeta F_B \big[ {\textstyle{0 \atop \beta+N/2}}\big] \right) \Bigg ]\, ,
\end{split} \label{eq:moebiusB}
\end{equation}
where
\begin{equation}
\begin{aligned}
   P_\beta^{(b)} &= \begin{cases}
       2^{b/2-4} \displaystyle{\sum_{\{ m \} }} \displaystyle{ \sum_{\{ \epsilon \in \{ 0,1 \} \} }}  \gamma_{\epsilon} \, q^{\frac{\alpha'}{2} (m+ \frac{1}{\alpha'} B \epsilon )^\intercal \, g^{-1} \, (m+ \frac{1}{\alpha'} B \epsilon ) } \equiv 2^{b/2-2}  P^{(\gamma)}(B,0) \, , \qquad &\beta=0 \, ,
\\
 \xi \, , \qquad &\beta \neq 0 \, ,
   \end{cases} 
   \end{aligned}
\end{equation}
and 
\begin{equation}
\begin{aligned}
W_\beta^{(b)} &= \begin{cases}  2^{-4/2} \displaystyle{ \sum_{\{n \} }} \displaystyle{  \sum_{\{ \epsilon \in \{ 0,1 \} \} }}  \tilde\gamma_{\epsilon}\, q^{\frac{1}{2\alpha'} n^\intercal \, g \, n} \,  e^{\frac{2 \pi i}{\alpha'} n^\intercal B \epsilon} \equiv W^{(\tilde{\gamma})}(0,B) \, , \qquad & \beta=0 \, ,
\\
\tilde \xi \, , \qquad &\beta \neq 0 \, .
   \end{cases} 
\end{aligned}
\end{equation}
A consistent particle interpretation requires the signs $\gamma$ and $\tilde \gamma$ to satisfy
\begin{equation}
     \sum_{\{ \epsilon  \in \text{ker}(B) \}} \gamma_\epsilon = \xi \cdot 2^{(4-b)/2} \, , \qquad \sum_{\{ \epsilon=0,1 \}} \tilde \gamma_\epsilon = \tilde{\xi} \cdot 2^{4/2} \, ,
\end{equation}
in addition to \eqref{eq:gammatr}. The signs $\xi$ and $\tilde \xi$ are those reported in \eqref{eq:transversemoebiusZN} and thus ascribe to the choice of the Wilson line or brane displacement. For example in the $T^4/\mathbb{Z}_4$ supersymmetric orientifold where all the O5$_-$ planes are placed on the $\mathbb{Z}_4$ fixed points, the Chan-Paton labels need to be complex and thus we have to enforce 
\begin{equation}
    \begin{aligned}
 &  \Lambda_1^{(b)} =  2^{(4-b)/2} \, , \qquad \tilde \Lambda_1^{(b)} = 1 \, .
\end{aligned}
\end{equation}
\noindent Before moving to the discussion of the concrete examples, let us just make one more comment on the nature of the orientifold projection according to the values of $\zeta$ and $\xi, \tilde \xi$. If $\zeta=1$, we have introduced D5 branes and bosons and fermions are both anti-symmetrised with a relative sign between the $T_B \big[ {\textstyle{0 \atop \beta}}\big]$ and the $T_B \big[ {\textstyle{0 \atop \beta+ N/2}}\big]$ blocks\footnote{This sign reflects the charges of O-planes and D-branes.}. If $\zeta=-1$ we have introduced $\overline{\text{D5}}$ branes, resulting in the absence of the relative sign between the relevant modular blocks and, most importantly, a different symmetrisation for bosons and fermions, which breaks supersymmetry.

\noindent As a final comment, we observe that, as anticipated, the values of $\xi$ and $\tilde \xi$ determine the parametrisation of the Chan-Paton charges that read
\begin{equation}
N_\beta = e^{ \pi i \beta /N} \sum_{\gamma=0}^{N-1} e^{2 \pi i \beta \gamma/N}\, n_\gamma\,,
\qquad
D_{(k)\,,\, \beta} = e^{ \pi i \beta /N} \sum_{\gamma=0}^{N-1} e^{2 \pi i \beta \gamma/N}\, d_{(k)\,,\,\gamma}\,,
\label{eq:CPcomplex}
\end{equation}
for $\xi, \tilde \xi=1$ and
\begin{equation}
N_\beta = \sum_{\gamma=0}^{N-1} e^{2 \pi i \beta \gamma/N}\, n_\gamma\,,
\qquad
D_{(k)\,,\, \beta} = \sum_{\gamma=0}^{N-1} e^{2 \pi i \beta \gamma/N}\, d_{(k)\,,\,\gamma}\,,
\label{eq:CPreal}
\end{equation}
for $\xi, \tilde \xi=-1$.

\noindent In the following, we will recall the construction for supersymmetric $\mathbb{Z}_2$ orbifold and we will complete the analysis in \cite{Angelantonj:1999jh} along the lines of \cite{Angelantonj:2024iwi}, where its BSB version has been discussed. We will therefore move on to the introduction of a rank $b$ Kalb-Ramond field for the other K3 orbifolds \cite{Aspinwall:1996mn} showing that, {\em algebraically}, no other consistent solution can be found except for the $\mathbb{Z}_4$ orientifold with a rank $2$ $B$-field turned on with O5$_-$ planes populating the $\mathbb{Z}_4$ fixed points.

\subsection{The supersymmetric \texorpdfstring{$T^4/\mathbb{Z}_2$}{T4Z2} orientifold}
\label{SSec:Z2B}

We shall begin our analysis by studying orientifold vacua built upon the $T^4/\mathbb{Z}_2$ orbifold. The new solutions for the BSB variation have already been discussed in \cite{Angelantonj:2024iwi} and thus we will focus only on the supersymmetric case. Following the general structure of the partition functions described in the previous Section, we can write the massless contribution for the Klein bottle amplitude with a non-vanishing $B$-field of rank $b$ in terms of combinations of characters $\tau^B_{\alpha,\beta}$, reported in the Appendix \ref{App:ZN}, forming eigenstates for the orbifold action as 
\begin{equation} \label{eq:kleinSUSYZ2B}
    \mathcal{K}^{(b)} = \tfrac{1}{2} (P+W (0,B) ) \tau^B_{0,0} + \tau^B_{0,1} + 16 \cdot 2^{-b/2} \tau^B_{1,0} \, ,
\end{equation}
where the modified multiplicity in the $g$-twisted sector $\Delta_{1,0}=16 \cdot 2^{-b/2}$ reproduces the net number of O-planes O5$_\pm$. The simultaneous presence of O5$_+$ and O5$_-$ planes induces different orientifold projections on closed string states, leading to  $4 + 8(1+ 2^{-b/2})$ neutral hyper multiplets and $1 + 8(1-  2^{-b/2})$ tensor multiplets. Performing an $S$ modular transformation allows to read the Klein bottle amplitude in the transverse channel 
\begin{equation} \label{eq:kleinZ2Btr}
    \tilde{\mathcal{K}}^{(b)} \sim \frac{2^{-5}}{2} \left \{ \tau^B_{0,0} \left (- 2^5 \sqrt{v} - \frac{2^{5-b/2}}{\sqrt{v}} \right )^2 +  \tau^B_{0,1} \left (- 2^5 \sqrt{v} + \frac{2^{5-b/2}}{\sqrt{v}} \right )^2 \right \} \, ,
\end{equation}
in which the reduced charge reflects the presence of sixteen orientifold planes of both types. 

\noindent The uncancelled R-R tadpole in \eqref{eq:kleinZ2Btr} then calls for the introduction of D9 and D5 branes to avoid anomalies. The contribution from the massless states in the transverse channel reads
\begin{equation}  \label{eq:annulusZ2Btr}
\begin{aligned}
    \tilde{ \mathcal{A}}^{(b)} \sim \frac{2^{-5}}{2} & \left \{ \tau^B_{0,0} \left ( 2^{b/2}\,  N_0 \sqrt{v} + \frac{1}{\sqrt{v}} \sum_i D_{(i),0} \right )^2 \right.
    \\
    & +  \tau^B_{0,1} \left ( 2^{b/2}\,  N_0 \sqrt{v} - \frac{1}{\sqrt{v}} \sum_i D_{(i),0} \right )^2
    \\
    & \left. + \frac{2^{b}}{2} \sum_{i=1}^{2^{4-b}} \tau^B_{1,0} \left ( N_1 - 4 \cdot 2^{-b/2} D_{(i),1} \right )^2 \right \} \, , 
    \end{aligned}
\end{equation}
when closed string states propagate between pairs of D9 and D5 branes, while it is given by
\begin{equation}  \label{eq:moebiusZ2Btr}
\begin{aligned}
    \tilde{\mathcal{M}}^{(b)} \sim - \frac{2}{2} & \left \{ \hat{\tau}^B_{0,0} \left ( 2^{b/2}\,  N_0 \sqrt{v} + \frac{1}{\sqrt{v}} \sum_i D_{(i),0} \right ) \left ( \sqrt{v} + \frac{2^{-b/2}}{\sqrt{v}} \right ) \right. 
    \\
    &\left. +  \hat{\tau}^B_{0,1} \left ( 2^{b/2}\,  N_0 \sqrt{v} - \frac{1}{\sqrt{v}} \sum_i D_{(i),0} \right ) \left ( \sqrt{v} - \frac{2^{-b/2}}{\sqrt{v}} \right ) \right \} \, ,
    \end{aligned}
\end{equation}
when closed string states propagate between D-branes and O-planes. Algebraically, the number of stacks of D5 branes to be introduced can be at most $2^{4-b}$, which corresponds to the only value allowing the desired factorisation of the R-R tadpoles \cite{Angelantonj:1999jh}. Notice that it does not correspond to the net number of fixed points and its geometrical interpretation is still elusive. Nevertheless, $2^{4-b}$ guarantees the consistency of the construction for an arbitrary rank of the Kalb-Ramond field $b=0,2,4$, where the $b=0$ reproduces the usual orientifold constructions in \cite{Bianchi:1990yu, Gimon:1996rq, Gimon:1996ay, Dabholkar:1996pc} and the other remaining values of $b$ reproduce other constructions already discussed in the literature \cite{Bianchi:1991eu, Gepner:1987qi, Angelantonj:1996mw}. 

\noindent The transverse channels in \eqref{eq:kleinZ2Btr}, \eqref{eq:annulusZ2Btr} and \eqref{eq:moebiusZ2Btr} encode the tadpole conditions 
\begin{equation}
    \begin{aligned}
        &2^{b/2} N_0= 2^5 \, , \qquad \sum_{i=1}^{2^{4-b}} D_{(i),0}= 2^{5-b/2} \, ,
        \\
        & N_1 - 4 \cdot 2^{-b/2} D_{(i),1}=0 \, , \qquad i=1, \ldots , 2^{4-b} \, .
    \end{aligned}
\end{equation}
where we have allowed D5 branes to populate all the available fixed points. The complete open spectrum is encoded into the vacuum amplitudes in eqs. \eqref{eq:annulus55B}, \eqref{eq:annulus99B}, \eqref{eq:annulus95B} and \eqref{eq:moebiusB}, where the structure of the gauge groups supported by D9 and D5 branes depends on the discrete Wilson lines and brane displacement introduced through $\xi$ and $\tilde \xi$\footnote{With abuse of terminology, we will interchangeably refer in the rest of the paper to the introduction of Wilson lines for D$9$ branes as brane displacement.}. Setting branes on a fixed point supporting an O5$_-$ plane induces a unitary gauge group while it is symplectic if the D$5$ branes are placed on a fixed point supporting an O5$_+$ plane, consistently with the analysis in \cite{Bianchi:1991eu}. The solutions exploited in \cite{Angelantonj:1999jh} only allow D9 and D5 branes to be placed on the same fixed point, supporting either an O5$_-$ by choosing $\xi=\tilde \xi=1$ or an O5$_+$ plane with $\xi=\tilde \xi=-1$. In the former case, the compatibility with the tadpole conditions identifies the only solution in \cite{Angelantonj:1999jh}, where the gauge group reads
\begin{equation} \label{eq:CPUUZ2B}
     G_{\text{CP}}= \text{U}(2^{4-b/2}) \times \prod_{i=1}^{2^{4-b}} \text{U}( 2^{b/2})
\end{equation}
and the light spectrum comprises a vector multiplet in the adjoint of \eqref{eq:CPUUZ2B}, one hyper multiplet in 
\begin{equation}
    ( \smalltableau{ \null \\ \null \\}+ \overline{ \smalltableau{ \null \\ \null \\}};1 ) + \sum_i ( 1; \smalltableau{ \null \\ \null \\}_i+ \overline{ \smalltableau{ \null \\ \null \\}}_i )  \, ,
\end{equation}
and  $2^{b/2}$ hyper multiplets in 
\begin{equation}
  \sum_i  ( \smalltableau{ \null \\} ; \overline{\smalltableau{ \null \\}}_i) + ( \overline{ \smalltableau{ \null \\}};  \smalltableau{ \null \\}_i )  \, .
\end{equation} 
The anomaly arising from this massless spectrum can be straightforwardly computed and reads
\begin{equation} \label{susy2pol8B1}
\begin{aligned}	
I_8 &= \tfrac{1}{64} \Bigg( -4(1+2^{-b/2}) \text{tr} R^2 + 2^{b/2+1} \, \text{tr} F_{9}^2 + 2 \,\sum_{i=1}^{2^{4-b}} \text{tr} F_{5_i}^2\Bigg)^2\\
 &- \tfrac{1}{64} \Bigg( -4(1-2^{-b/2}) \text{tr} R^2 + 2^{b/2+1}  \, \text{tr} F_{9}^2 - 2 \,\sum_{i=1}^{2^{4-b}} \text{tr} F_{5_i}^2\Bigg)^2 \, ,
 \end{aligned}
\end{equation}
reflecting the factorisation that allows the Green-Schwarz-Sagnotti mechanism \cite{Green:1984bx, Green:1984sg, Sagnotti:1992qw} to take place.

\noindent The choice $\xi=\tilde \xi=-1$ instead yields a product of symplectic gauge groups, not uniquely fixed by the tadpoles once D5 branes are placed over all the fixed points \cite{Aldazabal:1999nu, Angelantonj:2024iwi}. Indeed, the general solution gives the Chan-Paton gauge group
\begin{equation} \label{eq:CPUSpUSp2}
\begin{aligned}
    G_{\text{CP}}= & \text{USp}(2^{4-b/2} + 2a) \times \text{USp}(2^{4-b/2}- 2a) 
    \\
    &\times \prod_{i=1}^{2^{4-b}} \text{USp}( 2^{b/2-1}(2+a)) \times \text{USp}( 2^{b/2-1}(2-a)) \, , 
\end{aligned}\end{equation}
where $a=0, 2$ if $b=2$ and $a=0,1,2$ if $b=4$\footnote{Such solution cannot exist if $b=0$.}. Notice that the solution $a=0$ reproduces the standard one, already discussed in \cite{Angelantonj:1999jh}.  For this family of solutions, the massless spectrum comprises one vector multiplet in the adjoint of \eqref{eq:CPUSpUSp2}, one hyper multiplet in 
\begin{equation}
    ( \smalltableau{ \null \\}, \smalltableau{ \null \\} ; 1,1) + \sum_i ( 1, 1; \smalltableau{ \null \\}_i, \smalltableau{ \null \\}_i )  \, ,
\end{equation}
from string stretched between D9 branes and between D5 branes, and $2^{b/2-1}$ hyper multiplets in
\begin{equation}
  \sum_i  ( \smalltableau{ \null \\}, 1; 1, \smalltableau{ \null \\}_i) + ( 1,  \smalltableau{ \null \\} ; \smalltableau{ \null \\}_i )  \, ,
\end{equation}
from the D9/D5 amplitude\footnote{The extra degeneracy $2^{b/2}$ is needed to interpret D-branes as physical objects \cite{Angelantonj:1999jh} and can be ascribed to the torsional piece of the D9 gauge group \cite{Kakushadze:2000hm}.}. In such a case, the anomaly polynomial is given by
\begin{equation} \label{susy2pol8B4}
\begin{aligned}	
I_8 &= \tfrac{1}{64} \Bigg( -4(1+2^{-b/2}) \text{tr} R^2 + 2^{b/2}  \, \text{tr} F_{9,1}^2 + 2^{b/2}  \, \text{tr} F_{9,2}^2 -  2\,\sum_{i=1}^{2^{4-b}} \text{tr} F_{5_i}^2 \Bigg)^2
\\
 &- \tfrac{1}{64} \Bigg( -4(1-2^{-b/2}) \text{tr} R^2 + 2^{b/2}  \, \text{tr} F_{9,1}^2 + 2^{b/2}  \, \text{tr} F_{9,2}^2 -  2\,\sum_{i=1}^{2^{4-b}} \text{tr} F_{5_i}^2 \Bigg)^2  
 \\
 &- \tfrac{2^{b}}{128}  \sum_{k=1}^{2^{4-b}}  \Bigg( \text{tr} F_{9,1}^2 -  \text{tr} F_{9,2}^2 - 2^{2-b/2} \, \text{tr} F_{5_i,1}^2 + 2^{2-b/2} \,  \text{tr} F_{5_i,2}^2\Bigg)^2 \, ,
 \end{aligned}
\end{equation}
which again reproduces the factorised form allowing the Green-Schwarz-Sagnotti to take place.

\noindent The other two possibilities where D9 and D5 branes sit on fixed points supporting different O-planes have not been described in the literature so far. In this case, the gauge group is a product of unitary and symplectic ones and the orbifold projection term is absent \cite{Angelantonj:2024iwi} in the D9/D5 amplitude. In particular, choosing to place the D9(D5) branes in correspondence of O5$_-$(O5$_+$) planes\footnote{As explained in a previous footnote, moving D$9$ branes to a fixed point supporting an O$5_\pm$ plane means introducing a proper discrete Wilson line.} through the choice $\xi=-\tilde \xi=1$ yields
\begin{equation} \label{eq:CPUUSp2}
    G_{\text{CP}}= \text{U}(2^{4-b/2}) \times \prod_{i=1}^{2^{4-b}} \text{USp}(2^{b/2}) \times \text{USp}(2^{b/2}) \, .
\end{equation}
The light spectrum then comprises a vector in the adjoint of \eqref{eq:CPUUSp2}, one hyper multiplet in
\begin{equation}
    ( \smalltableau{ \null \\ \null \\}+ \overline{ \smalltableau{ \null \\ \null \\}}, 1 ; 1,1) + \sum_i (  1; \smalltableau{ \null \\}_i, \smalltableau{ \null \\}_i )  \, ,
\end{equation}
and $2^{b/2-1}$ hyper multiplets in 
\begin{equation}
  \sum_i  ( \smalltableau{ \null \\}; \smalltableau{ \null \\}_i, 1) + (  \smalltableau{ \null \\} ; 1, \smalltableau{ \null \\}_i )  \, .
\end{equation}
This model is anomaly-free as it can be shown by computing the anomaly polynomial
\begin{equation} \label{susy2pol8B2}
\begin{aligned}	
I_8 &= \tfrac{1}{64} \Bigg( -4(1+2^{-b/2}) \text{tr} R^2 + 2^{b/2+1} \, \text{tr} F_{9}^2 +  \sum_{i=1}^{2^{4-b}} \text{tr} F_{5_i,1}^2 +  \sum_{i=1}^{2^{4-b}} \text{tr} F_{5_i,2}^2\Bigg)^2
\\
 &- \tfrac{1}{64} \Bigg( -4(1-2^{-b/2}) \text{tr} R^2 + 2^{b/2+1}  \, \text{tr} F_{9}^2 -  \sum_{i=1}^{2^{4-b}} \text{tr} F_{5_i,1}^2 -  \sum_{i=1}^{2^{4-b}} \text{tr} F_{5_i,2}^2\Bigg)^2  
 \\
 &- \tfrac{1}{8}  \sum_{k=1}^{2^{4-b}}  \Bigg( \text{tr} F_{5_i,1}^2 -  \text{tr} F_{5_i,2}^2\Bigg)^2 \, .
 \end{aligned}
\end{equation}

\noindent Finally, a similar solution can be found with $\xi=-\tilde \xi=-1$
\begin{equation} \label{eq:CPUSpU2}
    G_{\text{CP}}= \text{USp}(2^{4-b/2}) \times \text{USp}(2^{4-b/2}) \times \prod_{i=1}^{2^{4-b}} \text{U}(2^{b/2}) \, ,
\end{equation}
where D9 branes are moved to a fixed point supporting an O5$_+$ plane and D5 branes are placed on a fixed point supporting an O5$_-$ plane. In such a case the light spectrum comprises a vector multiplet in the adjoint of \eqref{eq:CPUSpU2}, one hyper multiplet in 
\begin{equation}
    ( \smalltableau{ \null \\},   \smalltableau{ \null \\}; 1) + \sum_i ( 1, 1; \smalltableau{ \null \\ \null \\}_i+ \overline{ \smalltableau{ \null \\ \null \\}}_i )  \, ,
\end{equation}
and  $2^{b/2-1}$ hyper multiplets in 
\begin{equation}
  \sum_i  ( \smalltableau{ \null \\}, 1 ; \smalltableau{ \null \\}_i) + ( 1,   \smalltableau{ \null \\} ;  \smalltableau{ \null \\}_i )  \, .
\end{equation}
It is worth stressing that in both situations the presence of unitary groups implies that the R-R tadpoles unambiguously fix the gauge group, and the additional solutions can only be present for the \eqref{eq:CPUSpUSp2} case. The anomaly for such model is encoded into 
\begin{equation} \label{susy2pol8B3}
\begin{aligned}	
I_8 &= \tfrac{1}{64} \Bigg( -4(1+2^{-b/2}) \text{tr} R^2 + 2^{b/2}  \, \text{tr} F_{9,1}^2 + 2^{b/2}  \, \text{tr} F_{9,2}^2 -  2\,\sum_{i=1}^{2^{4-b}} \text{tr} F_{5_i}^2 \Bigg)^2
\\
 &- \tfrac{1}{64} \Bigg( -4(1-2^{-b/2}) \text{tr} R^2 + 2^{b/2}  \, \text{tr} F_{9,1}^2 + 2^{b/2}  \, \text{tr} F_{9,2}^2 -  2\,\sum_{i=1}^{2^{4-b}} \text{tr} F_{5_i}^2 \Bigg)^2  
 \\
 &- \tfrac{2^{b}}{128}  \sum_{i=1}^{2^{4-b}}  \Bigg( \text{tr} F_{9,1}^2 -  \text{tr} F_{9,2}^2\Bigg)^2 \, ,
 \end{aligned}
\end{equation}
which is again cancelled through the Green-Schwarz-Sagnotti mechanism.

\subsection{The supersymmetric \texorpdfstring{$T^4/\mathbb{Z}_4$}{T4Z4} orientifold}
\label{SSec:Z4BSUSY}

In the $T^4/\mathbb{Z}_4$ case, the structure of fixed points is richer when compared to the $T^4/\mathbb{Z}_2$ vacuum. Indeed, only four points are invariant under the orbifold action, while the other $\mathbb{Z}_2$ fixed points form $\mathbb{Z}_4$ doublets. The different nature of such fixed points is manifest when analysing the properties of the orientifold planes to be placed on them. Indeed, even though all the O5$_-$ planes are neutral with respect to the twisted R-R six-forms, the O$5_+$ planes placed on $\mathbb{Z}_4$ fixed points have a non-vanishing twisted charge and are thus {\em fractional}, while the twisted R-R charge vanishes for those placed on the $\mathbb{Z}_4$ doublets \cite{Angelantonj:2024iwi}. As discussed in \cite{Angelantonj:2024iwi}, the presence of a non-vanishing twisted charge highly constrains the structure and dynamics of the vacuum itself, since it requires the presence of $\overline{\text{D}5}$ branes to be stuck on the fixed points to neutralise the R-R charge locally. However, it turns out that the latter possibility is not compatible with a non-vanishing value for the $B$-field, which only allows O5$_-$ planes to populate the $\mathbb{Z}_4$ fixed points, as discussed in Section \ref{SSec:discussion}.

\noindent From eqs. \eqref{eq:ZNfixedpoints} and \eqref{eq:Z2fixedpoints}, placing the O5$_-$ planes on the $\mathbb{Z}_4$ fixed points allows to write down the massless Klein bottle amplitude
\begin{equation}\label{eq:kleinBSUSYZ4}
    \mathcal{K}^{(b)} =  \tfrac{1}{2} (P+W (0,B) ) \tau^B_{0,0} +  \left ( 4 \, - 2+ 8 \cdot 2^{-b/2} \right ) \tau^B_{2,0} \, ,
\end{equation}
where, following the conventions of Section \ref{Sec:6dorientifold}, $\Delta_{2,0}=16 \cdot 2^{-b/2}$ and $\Delta_{2,1}=4$. These values reflect the presence of $4$ O5$_-$ planes on the top of the $\mathbb{Z}_4$ fixed points, $4 + 8 \cdot 2^{-b/2}$ O5$_-$ planes and $8-8 \cdot 2^{-b/2}$ O5$_+$ planes on the $\mathbb{Z}_2$ fixed points forming $\mathbb{Z}_4$ doublets, giving rise to $9-4 \cdot 2^{-b/2}$ tensor and $12+ 4 \cdot 2^{-b/2}$ hyper multiplets. The charge and tension of the orientifold planes can be read in the transverse channel
\begin{equation}\label{eq:kleinBSUSYZ4tr}
\begin{aligned}
    \tilde{\mathcal{K}}^{(b)} =& \frac{2^{-5}}{4} \left ( - 2^5\, \sqrt{v} - \frac{2^{5-b/2}}{\sqrt{v}}\right)^2 \tau^B_{0,0} + \frac{2^{-5}}{4} \left ( - 2^5\, \sqrt{v} + \frac{2^{5-b/2}}{\sqrt{v}}\right)^2 (\tau^B_{0,1} +\tau^B_{0,3}) \, ,
\end{aligned}
\end{equation}
from which we see that all the orientifold planes do not carry charges under twisted R-R forms.

\noindent As before, the cancellation of anomalies implies that one has to introduce D9 and D5 branes with positive tension and charge, which cancels the R-R tadpoles. The massless contribution to the transverse channel for closed string states propagating between D-branes reads
\begin{equation} \label{eq:annulusBtrSUSYZ4}
\begin{split}
\tilde{\mathcal A}^{(b)} &= \frac{2^{-5}}{4}\left[ \left( 2^{b/2} \, N_0 \sqrt{v} + \frac{1}{\sqrt{v}} \sum_i D_{(i),0} \right)^2 \tau^B_{0,0} +\left( 2^{b/2} \, N_0 \sqrt{v} - \frac{1}{\sqrt{v}} \sum_i D_{(i),0} \right)^2 \left( \tau^B_{0,1} + \tau^B_{0,3} \right)\right.
\\
& \qquad \qquad +2  \cdot   2^b \sum_{\alpha =1,3}\sum_{i=1}^{2^{2-b}} (N_{\alpha} - 2 \cdot 2^{-b/2} D_{(i),\alpha})^2 \tau^B_{\alpha , 0}  
\\
& \left. \qquad \qquad + 12\, N_2^2  + 2^b \sum_{i=1}^{2^{2-b}} (N_{2} - 4 \cdot 2^{-b/2} D_{(i),2})^2 \tau^B_{2,0} \right] \,,
\end{split}
\end{equation}
and for closed strings propagating between D-branes and O-planes
\begin{equation} \label{eq:moebiusBtrSUSYZ4}
\begin{split}
\tilde{\mathcal M}^{(b)} =&-\frac{1}{2} \left( \sqrt{v} + \frac{2^{-b/2}}{\sqrt{v}}\right) \left( 2^{b/2} \, N_0 + \frac{1}{\sqrt{v}}\sum_{i} D_{(i),0}\right) \hat\tau^B_{0,0}+
\\
&-\frac{1}{2} \left(  \sqrt{v} - \frac{2^{-b/2}}{\sqrt{v}}\right) \left( 2^{b/2} \, N_0 - \frac{2^{-1}}{\sqrt{v}}\sum_{i} D_{(i),0}\right) \, (\hat\tau^B_{0,1}+ \hat\tau^B_{0,3}) \, .
\end{split}
\end{equation}
Notice that, algebraically, the number of stacks of branes is forced to be $2^{2-b}$ to factorise the tadpoles, thus implying that the only consistent results are provided by $b=0,2$. Furthermore, the structure of the tree-level amplitudes can only be consistent with a choice of Wilson lines and brane displacement enforcing complex Chan-Paton labels both for D9 and D5 branes, implying a vanishing contribution to the $g^2$-twisted sector in the M\"obius strip amplitude. Summing the contributions coming from $\mathcal{K}^{(b)}$, $\mathcal{A}^{(b)}$ and $\mathcal{M}^{(b)}$, we can obtain the tadpole conditions which, focusing on the $b=2$ case, reads
\begin{equation}
    \begin{aligned}
        &N_0= D_0= 16
        \\
        &N_{1,3} - D_{1,3}=0
        \\
        & N_2=0 \, , \qquad \text{and} \qquad D_2=0 \, .
    \end{aligned}
\end{equation}
Imposing the conditions above with the paramerisation of the Chan-Paton labels reported in eq. \eqref{eq:CPcomplex} leads to the gauge group 
\begin{equation}
    G_{\text{CP}}= \text{U}(4+a) \times  \text{U}(4-a) \Big|_{\text{D9}} \times \text{U}(4+a) \times  \text{U}(4-a) \Big|_{\text{D5}} 
\end{equation}
with $a=0,\ldots 4$, obtained from the Chan-Paton labels 
\begin{equation}
    n_0=n_3=4+a \, , \quad n_1=n_2=4-a \, , \quad d_{0}=d_{3}=4+a \, ,\quad d_{1}=d_{2}= 4-a \, .
\end{equation}
The massless spectrum follows from the direct channel partition functions described in eqs. \eqref{eq:annulus99B}, \eqref{eq:annulus55B} and \eqref{eq:moebiusB} for the D9-D9 and D5-D5 sectors, where the sum should be taken over one element since there is only one stack of branes for $b=2$. Furthermore, there is also the contribution from strings stretched between D9 and D5 branes
\begin{equation}
\begin{aligned}
    \mathcal{A}_{95}^{(2)}= 2 &  \left \{ (n_0 \, d_{3}  + n_1 \, d_{2} + n_2 \, d_{1} + n_3 \, d_{0}) \tau_{2,0} \right \} \, ,
\end{aligned}
\end{equation}
where the extra $2$ shows the presence of the extra degeneracy in eq. \eqref{eq:annulus95B}. The light spectrum thus comprises a vector multiplet in the adjoint of the gauge group, one hyper multiplet in
\begin{equation}
    ( \smalltableau{ \null \\ \null \\}, 1; 1, 1) + ( 1, \overline{\smalltableau{ \null \\ \null \\}}; 1, 1) + ( \smalltableau{ \null \\}, \overline{\smalltableau{ \null \\}}; 1, 1) \, ,
\end{equation}
from the D9/D9 amplitude, one hyper multiplet in 
\begin{equation}
(1,1; \smalltableau{ \null \\ \null \\}, 1) +  (1,1; 1, \overline{\smalltableau{ \null \\ \null \\}}) +  ( 1,1;\smalltableau{ \null \\}, \overline{\smalltableau{ \null \\}}) \, ,
\end{equation}
from the D5/D5 amplitudes and $2$ hyper multiplets in 
\begin{equation}
 (\smalltableau{ \null \\},1; \overline{\smalltableau{ \null \\}}, 1) + (1,\smalltableau{ \null \\};1, \overline{\smalltableau{ \null \\}})  \, ,
\end{equation}
from open strings stretched between D9 and D5 branes. The spectrum is free of local anomalies since the irreducible ones are cancelled and the reducible piece factorises as
\begin{equation} \label{eq:susy4pol8B}
	\begin{split}
		I_8 &= \tfrac{1}{128} \left( -6 \, \text{tr} R^2 + 4\, \text{tr} F_{9,1}^2 + 4\,  \text{tr} F_{9,2}^2 + 2 \, \text{tr} F_{5,1}^2 + 2 \, \text{tr} F_{5,2}^2 \right)^2
  \\
  &- \tfrac{1}{128} \left ( -2 \, \text{tr} R^2 + 4\, \text{tr} F_{9,1}^2 + 4\,  \text{tr} F_{9,2}^2 - 2\,  \text{tr} F_{5,1}^2 - 2 \, \text{tr} F_{5,2}^2 \right)^2
		\\
		& -\tfrac{4}{32}  \left( \text{tr} F_{9,1}^2 - \text{tr} F_{9,2}^2 -  \text{tr} F_{5,1}^2 + \text{tr} F_{5,2}^2 \right)^2 \,.
	\end{split}
\end{equation}

\subsection{The BSB \texorpdfstring{$T^4/\mathbb{Z}_4$}{T4/Z4} orientifold} \label{SSec:Z4BBSB}

\noindent Brane Supersymmetry Breaking (BSB) \cite{Antoniadis:1999xk, Mourad:2017rrl, Angelantonj:2024iwi} is a phenomenon occurring in lower dimensions in which a non-BPS configuration of BPS orientifold planes and D-branes induce a non-linear realisation of supersymmetry \cite{Dudas:2000bn, Pradisi:2001yv} without tachyonic instabilities. The simplest instance in which this situation occurs is the ten-dimensional Sugimoto model \cite{Sugimoto:1999tx}, where O9$_+$ planes are introduced at the place standard O9$_-$ planes occurring in the type I superstring. These objects carry positive tension and charge and thus call for the introduction of anti-branes, $\overline{\text{D9}}$ branes, to avoid anomalies. As a result, supersymmetry is broken in the open sector since gauge fields and {\em would be} gauginos transform in different representations of the $\text{USp}(32)$ gauge group. Although puzzling, the consistency of the coupling for the massless gravitino is guaranteed by noticing that supersymmetry is non-linearly realised \cite{Dudas:2000bn}. The main novelty introduced in BSB vacua \cite{Antoniadis:1999xk} is given by the presence of orientifold planes of different dimensionalities, O9 and O5 planes, so that it is possible to dress the world-sheet parity operator with an involution twisting charge and tension of only a subset of O-planes. The common choice in literature, O9$_-$ and O5$_+$ planes, unavoidably require D9 and $\overline{\text{D5}}$ branes to cancel anomalies. As before, supersymmetry is non-linearly realised \cite{Pradisi:2001yv}, since bosonic and fermionic open-string states transform into different representations, similarly to what happens in ten dimensions. This set-up has been realised for the first time in \cite{Antoniadis:1999xk} by taking the BSB orientifold of the $T^4/\mathbb{Z}_2$ orbifold and recently in \cite{Angelantonj:2024iwi} for the remaining orbifolds admitting a BSB variation: the $T^4/\mathbb{Z}_4$ and the $T^4/\mathbb{Z}_6$ orientifolds.

\noindent As emphasised in \cite{Angelantonj:2024iwi}, the BSB vacuum involving the $\mathbb{Z}_4$ orientifold induces a {\em truly} rigid configuration of branes, reflected in the absence of massless scalars transforming into bi-fundamental representations of the $\overline{\text{D5}}$ gauge group. Such rigidity is ascribed to the presence of fractional O5$_+$ planes on $\mathbb{Z}_4$ fixed points, forbidding the presence of the open string moduli which would allow a brane recombination. However, it is possible to deform such model by turning on non-vanishing Kalb-Ramond background field as done in \cite{Angelantonj:1999jh, Angelantonj:2024iwi} for the $T^4/\mathbb{Z}_2$ case, aiming to replace the fractional orientifold planes on the four $\mathbb{Z}_4$ fixed points with O5$_-$ planes with vanishing twisted charge.

\noindent We can start by discussing the case in which O5$_-$ planes are placed on the top of the $\mathbb{Z}_4$ fixed points. Following the general structure reported in section \ref{Sec:6dorientifold}, the Klein bottle amplitude reads
\begin{equation}\label{eq:kleinBSBZ4B}
    \mathcal{K} = \tfrac{1}{2} (P+W (0,B) )\tau^B_{0,0} +  \left ( 4 \, - 2-8 \cdot 2^{-b/2} \right ) \tau^B_{2,0} \, ,
\end{equation}
In eq. \eqref{eq:kleinBSBZ4B} it is possible to read that among the sixteen O5 planes, we have placed $4$ O5$_-$ planes on the $\mathbb{Z}_4$ fixed points and the remaining $4-8\cdot 2^{-b/2}$ O5$_-$ and $8+8\cdot 2^{-b/2}$ O5$_+$ on the $\mathbb{Z}_2$ fixed points forming $\mathbb{Z}_4$ doublets, thus giving rise to $12-4\cdot 2^{-b/2}$ uncharged hyper multiplets and $9 + 4 \cdot 2^{-b/2}$ tensor multiplets. Such configuration avoids the presence of fractional orientifold planes lying in the $\mathbb{Z}_4$ fixed points, implying the massless Klein bottle amplitude in the transverse channel to be
\begin{equation}\label{eq:kleinZ4Btr}
    \tilde{\mathcal{K}}^{(b)} = \frac{2^{-5}}{4} \left ( - 2^5\, \sqrt{v} + \frac{2^{5-b/2}}{\sqrt{v}}\right)^2 \tau^B_{0,0} + \frac{2^{-5}}{4} \left ( - 2^5\, \sqrt{v} - \frac{2^{5-b/2}}{\sqrt{v}}\right)^2 (\tau^B_{0,1} +\tau^B_{0,3}) \, ,
\end{equation}
where it is manifest that the orientifold planes carry only untwisted R-R charges.

\noindent Such uncancelled tadpoles require the introduction of D9 and $\overline{\text{D5}}$ branes\footnote{It is worth to notice that the presence of $\overline{\text{D5}}$ branes is dictated by the relative number of O5$_+$ and O5$_-$ planes forming $\mathbb{Z}_2$ doublets. If there are more O5$_+$ doublets than O5$_-$ ones, we are forced to introduce $\overline{\text{D5}}$ branes, while we are forced to introduce D5 branes in the opposite case.} to avoid anomalies. The contribution to the transverse channel of unoriented closed string states between D-branes is encoded into the transverse annulus partition function
\begin{equation} \label{eq:annulusZ4Btr}
\begin{aligned}
    \tilde{\mathcal{{A}}}^{(b)} = \frac{2^{-5}}{4} & \Bigg \{  \Big ( 2^{b/2} N_0 \sqrt{v} + \frac{1}{\sqrt{v}}\sum_{i=1}^{2^{2-b}} D_{(i),0} \Big )^2\, \left( \beta^B_{0,0} - \varphi^B_{0,1} - \varphi^B_{0,3} \right) 
    \\
    & +\Big ( 2^{b/2} N_0 \sqrt{v} - \frac{1}{\sqrt{v}}\sum_{i=1}^{2^{2-b}} D_{(i),0} \Big )^2\, \left( \beta^B_{0,1} - \beta^B_{0,3} - \varphi^B_{0,0} \right)
\\
& + 2 \cdot 2^{b} \sum_{i=1}^{2^{2-b}} \sum_{\alpha=1,3} \left [ \left (N_\alpha - 2 \cdot 2^{-b/2} D_{(i),\alpha}\right )^2 \beta^B_{\alpha , 0} \right.
\\
& \qquad \qquad \qquad \qquad \left.- \left (N_\alpha + 2 \cdot 2^{-b/2} D_{(i),\alpha}\right )^2 \varphi^B_{\alpha , 0} \right ]
\\
& + \Big [  2^{b} \sum_{i=1}^{2^{2-b}} \left( \left (N_2 - 4 \cdot 2^{-b/2} D_{(i),2} \right )^2 \beta^B_{2,0} \right.  
\\
& \qquad \qquad  \left. -\left (N_2 + 4 \cdot 2^{-b/2} D_{(i),2} \right )^2 \varphi^B_{2,0} \right) 
\\
&\qquad \qquad  + 12\, N_2^2 (\beta^B_{2,0}-\varphi^B_{2,0} ) \Big  ]  \Bigg \} \, ,  
\end{aligned}
\end{equation}
where we have decided to place $\overline{\text{D5}}$ brane only on the $\mathbb{Z}_4$ fixed points. The propagation of massless closed string states between D-branes and O-planes is instead encoded in the transverse channel of the M\"obius strip amplitude, which reads 
\begin{equation} \label{eq:moebiusZ4Btr}
\begin{aligned}
\tilde{\mathcal M}^{(b)} =-\frac{1}{2} & \Bigg \{ \left( \sqrt{v} - \frac{2^{-b/2}}{\sqrt{v}}\right) \left( 2^{b/2}N_0 + \frac{1}{\sqrt{v}}\sum_{i=1}^{2^{2-b}} D_{(i),0}\right) \, \hat\beta^B_{0,0}
\\
& - \left ( \sqrt{v} - \frac{2^{-b/2}}{\sqrt{v}} \right ) \left( 2^{b/2} N_0 - \frac{1}{\sqrt{v}}\sum_{i=1}^{2^{2-b}} D_{(i),0}\right) \, \hat\varphi^B_{0,0}
\\
&+ \left( \sqrt{v} + \frac{2^{-b/2}}{\sqrt{v}}\right) \left( 2^{b/2} N_0 - \frac{1}{\sqrt{v}}\sum_{i=1}^{2^{2-b}} D_{(i),0}\right) \, (\hat\beta^B_{0,1}+ \hat\beta^B_{0,3}) 
\\
& -  \left( \sqrt{v} + \frac{2^{-b/2}}{\sqrt{v}}\right) \left(2^{b/2} N_0 + \frac{1}{\sqrt{v}}\sum_{i=1}^{2^{2-b}} D_{(i),0}\right) \, (\hat\varphi^B_{0,1}+\hat\varphi^B_{0,3}) \Bigg \}\,.
\end{aligned}
\end{equation}
It is worth stressing that such configuration implies the Chan-Paton labels of D9 and $\overline{\text{D5}}$ branes to be complex, in order to guarantee a vanishing contribution in the $g^2$-twisted sector of the transverse M\"obius strip amplitude \eqref{eq:moebiusZ4Btr} and thus the factorisation of the twisted R-R tadpoles. Such a scenario is indeed consistent with D9 and $\overline{\text{D5}}$ branes being placed on the top of O5$_-$ planes supported by the $\mathbb{Z}_4$ fixed points. Moreover, as for the supersymmetric case, the configuration is consistent only if the rank of the background Kalb-Ramond field corresponds to $b=2$, since for $b=4$ the number of branes appearing in the $g^2$-twisted sectors in \eqref{eq:annulusZ4Btr} would be fractional. Gathering the amplitudes \eqref{eq:kleinZ4Btr}, \eqref{eq:annulusZ4Btr} and \eqref{eq:moebiusZ4Btr} together, we can read the tadpole conditions with $b=2$
\begin{equation}
\begin{aligned}
    &N_0= D_{0}= 16\, ,
    \\
    &N_1 +  \, D_{1}= 0 \,  ,
    \\
        & N_2=0 \, , \qquad \text{and} \qquad D_2=0 \, .
\end{aligned}
\end{equation}

\noindent In light of the considerations above, the Chan-Paton labels are complex and following \eqref{eq:CPcomplex} shall satisfy the tadpole conditions
\begin{equation}
\begin{aligned}
    &n_0=n_3=4+a \, , \qquad n_1=n_2= 4-a 
    \\ 
    &d_0=d_3=4-a \, , \qquad d_1=d_2=4+a \, ,
\end{aligned}
\end{equation}
with $a=0,\ldots, 4$. The gauge group is therefore
\begin{equation} \label{eq:CPBSB4}
    G_{\text{CP}}= \text{U}(4+a) \times  \text{U}(4-a) \Big|_{\text{D9}} \times \text{U}(4-a) \times  \text{U}(4+a) \Big|_{\overline{\text{D5}}} \, .
\end{equation}

\noindent The massless contribution to the D9/D9 annulus amplitude then reads
\begin{equation} \label{eq:annulus9Z4B}
    \mathcal{A}_{99}^{(2)}= (2 n_0 \, n_3 + 2 n_1 \, n_2)\, \tau^B_{0,0} +  (2 n_0 \, n_2 + n_1^2 + n_3^2 ) \tau^B_{0,1} +  (n_0^2 + n_2^2 +2  n_1 \, n_3) \, \tau^B_{0,3} \, ,
\end{equation}
while its orientifold projection is described by
\begin{equation}
{\mathcal M}_9^{(2)} = - (n_1 + n_3) \, \hat\tau^B_{0,1} - (n_0 + n_2) \, \hat\tau^B_{0,3}\, .
\label{eq:moebius9Z4B}
\end{equation}
Open strings stretched between $\overline{\text{D5}}$ branes are given by
\begin{equation} \label{eq:annulus5bZ4B}
    \mathcal{A}_{\bar{5} \bar{5}}^{(2)}= (2 d_0 \, d_3 + 2 d_1 \, d_2)\, \tau^B_{0,0}  +  (2 d_0 \, d_2 + d_1^2 + d_3^2 ) \tau^B_{0,1} +  (d_0^2 + d_2^2 + 2 d_1 \, d_3) \, \tau^B_{0,3}\, ,
\end{equation}
with orientifold projection 
\begin{equation}
{\mathcal M}_{\bar{5}}^{(2)} = - (d_1 + d_3) \, \left ( \hat\beta^B_{0,1} + \hat\varphi^B_{0,1} \right ) - (d_0 + d_2) \, \left ( \hat\beta^B_{0,3} + \hat\varphi^B_{0,3} \right )\, .
\label{eq:moebius5bZ4B}
\end{equation}
Finally strings stretched between D9 and $\overline{\text{D5}}$ branes contribute to the massless spectrum via
\begin{equation}
\begin{aligned}
    \mathcal{A}_{9 \overline{5}}^{(2)}  = 2 & \left[ (n_0 \, d_{3}+n_2 \, d_{1}+ n_1 \, d_2 + n_3 \, d_0) \tau^{\hat B}_{2,0} \right.
    \\
    & + (n_0 \, d_{2}+n_1 \, d_{1}+ n_2 \, d_0 + n_3 \, d_3) \tau^{\hat B}_{2,1}
    \\
    & \left. + (n_0 \, d_{0}+n_1 \, d_{3}+ n_2 \, d_2 + n_3 \, d_1) \tau^{\hat B}_{2,3} \right]\, ,
    \end{aligned}\label{A95bZ4B}
\end{equation}
where again the multiplicity $2$ denotes the usual extra degeneracy.

\noindent As a result, the light spectrum comprises a gauge boson in the adjoint of \eqref{eq:CPBSB4}
a left-handed Weyl fermion in the representations
\begin{equation}
(\smalltableau{  \null \\} \times \overline{\smalltableau{ \null \\ }},1;1,1)+(1,\smalltableau{  \null \\} \times \overline{\smalltableau{ \null \\ }};1,1)+(1,1;\smalltableau{  \null \\} \times \overline{\smalltableau{ \null \\ }},1)+ (1,1;1,\smalltableau{  \null \\} \times \overline{\smalltableau{ \null \\ }})\,,
\end{equation}
four scalars in the representations
\begin{equation}
(\smalltableau{  \null \\ \null \\} , 1;1,1)+ (1, \overline{\smalltableau{  \null \\ \null \\}};1,1) + (\smalltableau{  \null \\}, \overline{\smalltableau{ \null \\}};1,1) +(1,1;\smalltableau{  \null \\ \null \\} ,1)+ (1, 1;1,\overline{\smalltableau{  \null \\ \null \\}}) + (1,1;\smalltableau{  \null \\}, \overline{\smalltableau{ \null \\}}) \,,
\end{equation}
a right-handed Weyl fermion in the representations
\begin{equation}
(\smalltableau{  \null \& \null \\} , 1;1,1)+ (1, \overline{\smalltableau{  \null \& \null \\}};1,1) + (\smalltableau{  \null \\}, \overline{\smalltableau{ \null \\}};1,1) +(1,1;\smalltableau{  \null \& \null \\} ,1)+ (1, 1;1,\overline{\smalltableau{  \null \& \null \\}}) + (1,1;\smalltableau{  \null \\}, \overline{\smalltableau{ \null \\}}) \,,
\end{equation}
two left-handed Weyl fermions in the representations 
\begin{equation}
 (\smalltableau{  \null \\ } ,  1; \overline{\smalltableau{  \null \\ }} , 1)+(1, \smalltableau{  \null \\ } ; 1, \overline{\smalltableau{  \null \\ }} )\,,
\end{equation}
and four real scalars in the representations
\begin{equation}
(\smalltableau{  \null \\ } ,1; \smalltableau{  \null \\ },1)+(1,\smalltableau{  \null \\ } ;\overline{\smalltableau{  \null \\ }},1 ) +(\overline{\smalltableau{  \null \\ }} ,1; 1,\smalltableau{  \null \\ })+(1,\overline{\smalltableau{  \null \\ }};1,\overline{\smalltableau{  \null \\ }} ) \,.
\end{equation}  
The model obtained in this way can be shown to be free of local anomalies since the anomaly polynomial factorises in such a way to allow the generalised Green-Schwarz-Sagnotti mechanism to take place
\begin{equation} \label{eq:bsb4pol8B}
	\begin{split}
		I_8 &= \tfrac{1}{128} \left( -2 \, \text{tr} R^2 + 4\, \text{tr} F_{9,1}^2 + 4\,  \text{tr} F_{9,2}^2 - 2 \, \text{tr} F_{5,1}^2 - 2 \, \text{tr} F_{5,2}^2 \right)^2
  \\
  &- \tfrac{1}{128} \left ( -6 \, \text{tr} R^2 + 4\, \text{tr} F_{9,1}^2 + 4\,  \text{tr} F_{9,2}^2 + 2\,  \text{tr} F_{\bar 5,1}^2 + 2 \, \text{tr} F_{\bar 5,2}^2 \right)^2
		\\
		& -\tfrac{4}{32}  \left( \text{tr} F_{9,1}^2 - \text{tr} F_{9,2}^2 +  \text{tr} F_{\bar 5,1}^2 - \text{tr} F_{\bar 5,2}^2 \right)^2 \,.
	\end{split}
\end{equation}

\noindent As expected, the configuration in such a case is less rigid and indeed scalars in bi-fundamental representations of $\overline{\text{D5}}$ gauge groups appear, showing how such branes can be further recombined via Higgsing. The minimal solution compatible with the tadpole identifies the gauge group
\begin{equation}
    G_{\text{CP}}=\text{SO}(2-a/2) \Big|_{\text{D9} \ \text{bulk}} \times \text{USp}(2-a/2) \Big|_{\overline{\text{D5}} \ \text{bulk}} \times \text{U}(4+a) \Big|_{\text{D9}}  \times \text{U}(4+a) \Big|_{\overline{\text{D5}}} \, ,
\end{equation}
with $a=0,2$.

\subsection{Discussion}
\label{SSec:discussion}

The extension of the analysis to the $T^4/\mathbb{Z}_6$ orbifold or to a rank $4$ Kalb-Ramond background field for the $T^4/\mathbb{Z}_4$ is obstructed by a common issue. Indeed, factorising the tadpoles in the transverse channel would require a fractional number of D-branes, as can be explicitly seen from eq. \eqref{eq:transverseannulusZN}. Indeed, the D-branes charge encoded into the  R-R tadpoles in the $g$ and $g^3$-twisted sectors in the $T^4/\mathbb{Z}_4$ orientifolds \eqref{eq:annulusBtrSUSYZ4} and \eqref{eq:annulusZ4Btr} read schematically
\begin{equation}
2^b \sum_{i=1}^{2^{2-b}} (N_{\alpha} \pm 2 \cdot 2^{-b/2} D_{(i),\alpha})^2 \, , \qquad \alpha=1,3 \, ,
\end{equation}
and thus a rank $b=4$ would imply a fractional number of D-branes. A similar result applies for the $T^4/\mathbb{Z}_6$ case, where the R-R tadpoles in the $g$ and $g^5$ twisted sectors read
\begin{equation}
  N_{\alpha}^2 + \sum_i D_{(i),\alpha}^2 \pm 2 \cdot 2^{b/2} \sum_i N_{\alpha} D_{(i),\alpha} =  2^b \sum_{i=1}^{2^{-b}} (N_{\alpha} \pm  2^{-b/2} D_{(i),\alpha})^2 \, , \qquad \alpha=1,5 \, ,
\end{equation}
which leads to a fractional number of D-branes for any $b$. A fractional number of D-branes has no physical meaning and thus the corresponding vacua would be inconsistent and hence anomalous\footnote{In \cite{Bachas:2008jv}, it is shown that introducing a $B$-field leads naturally to the presence of semi-integer charges admitting a T-dual geometric interpretation. In the present case, however, it is unclear if this consideration can be extended to higher-order orbifolds and how.}. Moreover, if such solutions are forbidden it would be interesting to understand whether or not this impossibility originates from topological considerations. Indeed, in toroidal compactification, the rank of the Kalb-Ramond field is interpreted as a characteristic class lying in $\mathbb{Z}_2$ cohomology encoding the obstruction for the gauge bundle to admit a vector structure \cite{Witten:1997bs}. On orbifolds, the second generalised Stiefel-Whitney class should be considered as an element of the $\mathbb{Z}_N$ equivariant cohomology of the torus $T^4$ and may imply non-trivial consequences\footnote{I would like to thank Ivano Basile for discussions on this point.}. In particular, if such obstructions are topological in nature, the $\mathbb{Z}_4$-twist should eliminate the maximal rank for the $\mathbb{Z}_2$ anti-symmetric tensor classifying bundles without vector structure, while the $\mathbb{Z}_6$ twist should allow only a trivial class. 

\noindent In addition, in the remaining $T^4/\mathbb{Z}_4$ with a rank $2$ $B$-field, although one would expect to yield a consistent vacuum for every configuration of orientifold planes, this is not true in general. Indeed, placing four O5$_+$ planes or two O5$_-$ and two O5$_+$ planes on the $\mathbb{Z}_4$ fixed points for the $T^4/\mathbb{Z}_4$ orbifold does not lead to a correct factorisation of the tadpoles when D-branes are added. Indeed the presence of a non-trivial background value for the Kalb-Ramond field induces a smaller number of D5 or $\overline{\text{D}5}$ branes which allows the cancellation of the charge associated to the non-dynamical ten form in ten dimensions but are not enough to compensate the contribution from all the six-forms localised on the various $\mathbb{Z}_4$ fixed points, whose charges need to be neutralised locally. Although this picture arises from algebraic considerations, its geometrical interpretation remains obscure and it would be interesting to carry out a deeper inspection to eventually corroborate this picture.   

\section{Unitarity constraints on string defects}\label{Sec:defects}

As described in the previous Sections, the light spectra give rise to an anomaly polynomial which has the general factorised form
\begin{equation}
    I_8= \frac12 \Omega_{\alpha \beta} X_4^\alpha X_4^\beta \, ,
\end{equation}
where $\Omega_{\alpha \beta}$ is a $\text{SO}(1,n_T)$ invariant metric and $X_4^\alpha$ is a $4$-form polynomial in the curvature of the gauge and tangent bundles
\begin{equation} \label{eq:tHooft}
    X_4^\alpha= \frac{1}{2} a^\alpha \text{tr}R^2 + \frac{1}{2} \sum_i \frac{b_i^\alpha}{\lambda_i} \text{tr}F_i^2 \, ,
\end{equation} 
with $\lambda_i$ the usual group factors denoting half of the length of the highest root and $a, \, b_i$ the 't Hooft coefficients. The shape of such polynomial follows from the structure of the tadpoles, which allows to cancel the irreducible anomalies \cite{Aldazabal:1999nu, Bianchi:2000de} and determines the structure of the Chern-Simons coupling in the low energy effective action for D-branes \cite{Douglas:1995bn, Green:1996dd} and O-planes \cite{Morales:1998ux, Stefanski:1998he}. As a result, the anomaly is cancelled by requiring an anomalous variation of the $2$-form fields via the celebrated Green-Schwarz-Sagnotti mechanism \cite{Green:1984bx, Green:1984sg, Sagnotti:1992qw}. However, this induces an anomaly on the world-volume of string defects coupled to the R-R $2$-forms, namely D1 branes localised in the internal manifold, or D5' branes which wrap all the internal cycles\footnote{The associated anti-branes are also allowed, but we will ignore this possibility.}. These defects are required to cancel the inflow from the bulk theory, which constrains the string defect charges $Q$. In addition, these have to satisfy the positivity condition for the tension of the defect
\begin{equation}
    Q \cdot J >0 \, ,
\end{equation}
where $J$ is the so-called {\em K\"ahler form} and is given by a combination of scalars parametrising the tensor branch of the moduli space of six-dimensional vacua. This is constrained by requiring the kinetic terms of scalars and gauge fields and the Gauss-Bonnet terms to be positive definite
\begin{equation} \label{eq:Jform}
    J \cdot J >0 \, , \qquad  J \cdot \Tilde{b}_i <0 \, , \qquad  J \cdot \tilde{a} >0 \, ,
\end{equation}
where in our setting $\tilde{b}_i$ and $\tilde{a}$ are encoded in the one-point functions of D-branes and O-planes with respect to the dilaton and the remaining (un)twisted NS-NS scalars to be read from the NS-NS tadpoles \cite{Angelantonj:2024iwi}. For the supersymmetric models, the R-R and NS-NS tadpoles are identical and therefore $\tilde{b}_i$ and $\tilde{a}$ are the 't Hooft coefficients in \eqref{eq:tHooft}, reproducing the original formulation of \cite{Sagnotti:1992qw, Kim:2019vuc}, but for BSB models $\tilde{a}=a$ and $\tilde{b}_i=-b_i$\footnote{In \cite{Angelantonj:2020pyr}, it has been shown how using the 't Hooft coefficients to the define $J$ also for BSB vacua would lead to the presence of ghosts.}.    
Once such defects are found, we can proceed by analysing the unitarity conditions along the lines of \cite{Angelantonj:2020pyr, Angelantonj:2024iwi}. Indeed, in the IR the massless degrees of freedom give rise to a two-dimensional CFT both on the left and right moving sectors with total central charges $c_{\text{L}}$ and $c_{\text{R}}$ \cite{Kim:2019vuc, Angelantonj:2024iwi}. On the other hand, the bulk decouples and the D9 and D5 or $\overline{\text{D5}}$ branes gauge groups become global symmetry realising in the IR a Ka$\check{\text{c}}$-Moody algebra both on the left or right moving sectors, whose levels $k_i$ can be extracted from the anomaly polynomial. This situation was already encountered in \cite{Angelantonj:2024iwi} when the D$p$ defect can be interpreted as an instanton of the gauge theory living on the D$(p+4)$ brane for BSB vacua. This is not however an intrinsic feature of such non-supersymmetric models but applies as well to the cases in which supersymmetry is linearly realised.   
Therefore, for the theory to be unitary we shall require\footnote{As emphasised in \cite{Angelantonj:2024iwi}, these conditions are {\em minimal} since the exact formulation would require the knowledge of the CFT realised by the bosons.}
\begin{equation} \label{eq:consistencycond}
    \sum_{i \, | \, k_i>0} \frac{k_i \text{dim}G_i}{k_i + h^\vee_i} \leq c_{\text{L}} -4_{\text{CM}}\, , \qquad \sum_{i \, | \, k_i<0} \frac{|k_i| \text{dim}G_i}{|k_i| + h^\vee_i} \leq c_{\text{R}}-6_{\text{CM}} \, ,
\end{equation}
where  $h^\vee_i$ denotes the dual Coxeter number associated to the $i$-th Ka$\check{\text{c}}$-Moody algebra and we have subtracted the contribution from the centre of mass which decouples in the IR.

\noindent In the following, we will show how the models at hand satisfy the consistency conditions in eq. \eqref{eq:consistencycond}, both in the case of non-trivial discrete Wilson line and brane displacement. The light spectrum of D1 and D5' branes can be straightforwardly computed in string perturbation theory following the rules in \cite{Dudas:2001wd} with some additional subtleties. Indeed, as for the D5 or $\overline{\text{D5}}$ branes, the D1 branes have Dirichlet boundary conditions along the internal compact directions, thus implying the D1/D9 amplitudes to have an extra degeneracy $2^{b/2}$. Similarly, the D5' branes have Neumann boundary conditions in the internal manifolds and therefore the D5'/D5 or $\overline{\text{D5}}$ amplitudes should possess the same extra degeneracy, following the D9/D5 or $\overline{\text{D5}}$ amplitudes. Moreover, as for the bulk theory, it is possible to turn on discrete Wilson lines for D5' branes or move the D1 branes to fixed points that can support O5$_-$ or O5$_+$ planes. In the former case, the Chan-Paton factors must be complex, while being real in the latter case. As a result, this choice unavoidably modifies the light spectrum, providing us different situations to be tested. In particular, we will focus on the $T^4/\mathbb{Z}_2$ supersymmetric vacuum with $\text{U}(2^{4-b/2})\times \text{USp}(2^{4-b/2}) \times \text{USp}(2^{4-b/2})$ gauge group and on the $T^4/\mathbb{Z}_4$ BSB orientifold yielding the $\text{U}(4) \times \text{U}(4) \times  \text{U}(4) \times \text{U}(4)$ gauge group. For the former, we will discuss D1 branes on the top of an O5$_-$ plane supporting a $ \text{U}(r)$ gauge group and D1 branes on the top of an O5$_+$ plane supporting a $\text{SO}(r_1) \times \text{SO}(r_2)$ gauge group, before discussing D5' branes yielding the gauge group $\text{USp}(r_1) \times \text{USp}(r_2)$. Afterwards, we will move to the BSB case, in which minimal D1 branes and D5' branes carrying a $\text{U}(r_1) \times \text{U}(r_2)$ gauge group are discussed.

\subsection{Examples}
\label{SSec:Examplesdefects}

We can start discussing the hybrid supersymmetric example described in Section \ref{SSec:Z2B} carrying $\text{U}(2^{4-b/2})\times \text{USp}(2^{4-b/2}) \times \text{USp}(2^{4-b/2})$ as a gauge group. The bulk theory identifies an anomaly lattice spanned by
\begin{equation} \label{susy2coeffB2}
	\begin{aligned}
		&a=\left ( -\sqrt{2}(1+2^{-b/2}),-\sqrt{2}(1-2^{-b/2}); \boldsymbol{0}_{n_T-1}\right ) \, ,
		\\
		&b_1= \left (\tfrac{2^{b/2}}{\sqrt{2}}, \tfrac{2^{b/2}}{\sqrt{2}} ; \boldsymbol{0}_{n_T-1}\right ) \, ,
		\\
		&b_{2}= \left (\tfrac{1}{2\sqrt{2}}, -\tfrac{1}{2\sqrt{2}}; \boldsymbol{\delta^1}_{2^{4-b}}; \boldsymbol{0}_{n_T-2^{4-b}-1} \right ) \, ,
       \\
        &b_{3}= \left (\tfrac{1}{2\sqrt{2}}, -\tfrac{1}{2\sqrt{2}};-\boldsymbol{\delta^1}_{2^{4-b}}; \boldsymbol{0}_{n_T-2^{4-b}-1} \right ) \, ,
	\end{aligned}
\end{equation}
where  we have introduced the $d$-dimensional vectors $\boldsymbol{\delta}^i_d$ with components $( \boldsymbol{\delta}^i_d )_j= \delta_j^i$ and $\boldsymbol{0}_d$ with components $( \boldsymbol{0}_d )_j= 0$.
In such a case, placing D1 branes on the top of an O5$_-$ plane implies a unitary gauge group $\text{U}(r)$, yielding the light spectrum reported in table \ref{tab:D1SUSYZ2UUSpB-} obtained using the definitions of Appendix B of \cite{Angelantonj:2024iwi}.   
\begin{table}
\centering
\begin{tabular}{| c | c | c |}
\hline
	 { Representation} &  $\text{SO}(1,1) \times \text{SU}(2)_u \times \text{SU}(2)_v $ & { Sector} \\
	\hline
	$ \left (\smalltableau{ \null \\} \times \overline{\smalltableau{\null \\}} \right )_1  $  & $(0, 1, 1 ) + 2 \times (\tfrac{1}{2}, 2, 1 )_L  $  & D1-D1
	\\
	$  \left (\smalltableau{ \null \\} \times \overline{\smalltableau{\null \\}} \right )_1  $ & $(1, 2, 2 ) + 2 \times (\tfrac{1}{2}, 1, 2 )_R$ & \ \ 
	\\
	$ \smalltableau{ \null \& \null \\}_1 + \overline{\smalltableau{ \null \& \null \\}}_1 $  & $4  (1, 1, 1 ) + 2 \times (\frac{1}{2}, 2, 1 )_R $ &  \ \ 
	\\
	 $  \smalltableau{ \null \\ \null \\}_1 + \overline{\smalltableau{ \null \\ \null \\}}_1 $  & $ 2 \times (\frac{1}{2},  1,2 )_L $ &  \ \
	\\ 
 \hline
	$2^{b/2}  (\, \smalltableau{\null \\}_{1}, \overline{\smalltableau{\null \\}}_{9} ) + 2^{b/2} (\,   \overline{\smalltableau{\null \\}}_{1} \, , \smalltableau{\null \\}_{9}) $  & $  (\frac{1}{2},  1, 1 )_L $ & D1-D9 
 \\
 \hline
\end{tabular}
\caption{ Spectrum for probe D1 branes on the $\mathbb{Z}_2$ fixed point supporting an O5$_-$ plane, yielding a $\text{U}(r)$ gauge group.}
\label{tab:D1SUSYZ2UUSpB-}
\end{table}
The anomaly polynomial arising from the microscopic theory 
\begin{equation} \label{eq:anomalypolI41}
    I_4= \tfrac{1}{2} \left ( -2 r \, \text{tr} R^2 + 2^{b/2} r \,\text{tr} F_9^2 \right )
\end{equation}
is cancelled by the inflow of the bulk theory if the charge reads
\begin{equation}
    Q= \left ( \tfrac{r}{\sqrt{2}}, - \tfrac{r}{\sqrt{2}}; \boldsymbol{0}_{n_T-1} \right ) \, . 
\end{equation}
From eq. \eqref{eq:anomalypolI41}, the level of the Ka$\check{\text{c}}$-Moody algebra reads
\begin{equation} \label{eq:kacmoodyUxUxUsp}
    k_1= 2^{b/2} \, r \, ,
\end{equation}
which enter the unitarity conditions \eqref{eq:consistencycond}. For the minimal choice $r=1$, they read
\begin{equation} \label{eq:SUSYZ2UUSpUsp}
\begin{split}
   & \sum_{i \, | \, k_i \geq 0} \frac{k_i \ \text{dim} G_i }{k_i + h_i^{\vee}}= \frac{2^{8-b}-1}{1+2^{4-b}}+1 < c_\text{L}-4_{\text{CM}}=26 \, , 
    \\
   & \sum_{i \, | \, k_i <0} \frac{|k_i| \ \text{dim} G_i }{|k_i| + h_i^{\vee}}=0 < c_\text{R}-6_{\text{CM}}=12 \, .
\end{split}
\end{equation}

\noindent When placing D1 branes on the top of an O5$_+$ plane, the parametrisation of the Chan-Paton factors supports a gauge group $\text{SO}(r_1) \times \text{SO}(r_2) $. The light spectrum is then modified and it is reported in table \ref{tab:D1SUSYZ2UUSpB+}. 
\begin{table}
\centering
\begin{tabular}{| c | c | c |}
\hline
	 { Representation} &  $\text{SO}(1,1) \times \text{SU}(2)_u \times \text{SU}(2)_v $ & { Sector} \\
	\hline
	$   \smalltableau{ \null \\ \null \\}_{1_1} +  \smalltableau{ \null \\ \null \\}_{1_2} $  & $(0, 1, 1 ) + 2 \times (\tfrac{1}{2}, 2, 1 )_L  $  & D1-D1
	\\
	$   \smalltableau{ \null \& \null \\}_{1_1} +  \smalltableau{ \null \& \null \\}_{1_2}  $ & $(1, 2, 2 ) + 2 \times (\tfrac{1}{2}, 1, 2 )_R$ & \ \ 
	\\
	$ ( \,  \smalltableau{ \null \\}_{1_1} ,  \smalltableau{ \null \\}_{1_2}) $  & $4 \times (1, 1, 1 ) + 2 \times (\frac{1}{2}, 2, 1 )_R $ &  \ \ 
	\\
	 $   ( \,  \smalltableau{ \null \\}_{1_1} ,  \smalltableau{ \null \\}_{1_2})  $  & $ 2 \times (\frac{1}{2},  1,2 )_L $ &  \ \
	\\ 
 \hline
	$2^{b/2}  \,  ( \,  \smalltableau{ \null \\}_{1_1} \, + \, \smalltableau{ \null \\}_{1_2} , \smalltableau{ \null \\}_{9}) $  & $  (\frac{1}{2},  1, 1 )_L $ & D1-D9 
 \\
 \hline
	$( \,  \smalltableau{ \null \\}_{1_1} ,  \smalltableau{ \null \\}_{\bar 5_1})   + ( \,  \smalltableau{ \null \\}_{1_2} ,  \smalltableau{ \null \\}_{\bar 5_2}) $  & $ (1, 2, 1 ) + 2 \times (\frac{1}{2},  1, 1 )_R $ & D1-D5 
 \\
	$( \,  \smalltableau{ \null \\}_{1_1} ,  \smalltableau{ \null \\}_{ 5_2})   + ( \,  \smalltableau{ \null \\}_{1_2} ,  \smalltableau{ \null \\}_{5_1}) $  & $ 2 \times (\frac{1}{2},  1, 1 )_L $ & \ \
 \\
 \hline
\end{tabular}
\caption{ Spectrum for probe D1 branes on the $\mathbb{Z}_2$ fixed point supporting an O5$_+$ plane, yielding a $\text{SO}(r_1) \times \text{SO}(r_2)$ gauge group.}
\label{tab:D1SUSYZ2UUSpB+}
\end{table}
As a result, the contribution to the anomaly polynomial of the microscopic theory is slightly different 
\begin{equation}
\begin{split}
    I_4= \tfrac{1}{2} & \left ( - (r_1 + r_2) \, \text{tr} R^2  - (r_1-r_2)^2 \chi(N)  \right.
    \\
    & \left. + 2^{b/2} \tfrac{r_1+r_2}{2} \,\text{tr} F_{9}^2 - (r_1-r_2)\,\text{tr} F_{ 5,1}^2 - (r_2-r_1)\,\text{tr} F_{5,2}^2 \right ) \, ,
\end{split}
\end{equation}
and it is cancelled by the bulk theory inflow if the charge reads
\begin{equation} \label{eq:defcharegSO}
    Q= \left ( \tfrac{1}{2\sqrt{2}}(r_1+r_2), - \tfrac{1}{2\sqrt{2}}(r_1+r_2); -r_2 + r_1; \boldsymbol{0}_{n_T-2} \right ) \, .
\end{equation}
We can therefore proceed as before in the evaluation of the unitarity constraints, by reading the gauge algebras 
\begin{equation}
    k_1= 2^{b/2} \cdot  \tfrac{r_1+r_2}{2} \, , \qquad  k_2= r_2-r_1 \, , \qquad \qquad  k_3= r_1-r_2 \, .
\end{equation}
For instance, we can consider the choice $r_1=1, \, r_2=0$\footnote{The opposite choice for $r_1$ and $r_2$ yields equivalent results.} which implies
\begin{equation}
    \begin{split}
    &\sum_{i \, | \, k_i \geq 0} \frac{k_i \ \text{dim} G_i }{k_i + h_i^{\vee}}= \frac{2^{8-b}-1}{2+2^{5-b}}+1 + \frac{2^{4-b/2}(2^{4-b/2}+1)}{4 + 2^{4-b/2}} \leq c_{\text{L}}-4_{\text{CM}}=8 + 48 \cdot 2^{-b/2} \, ,
    \\
    &\sum_{i \, | \, k_i <0} \frac{|k_i| \ \text{dim} G_i }{|k_i| + h_i^{\vee}}= \frac{2^{4-b/2}(2^{4-b/2}+1)}{4 + 2^{4-b/2}} \leq c_{\text{R}}-6_{\text{CM}}=48 \cdot 2^{-b/2} \, . 
\end{split}
\end{equation}

\noindent A similar discussion can be carried out when D5' branes are introduced. Choosing to consider a the discrete Wilson line which gives a real parameterisation of the Chan-Paton factors, implying $\text{USp}(r_1) \times \text{USp}(r_2) $ as a gauge group. As before, one can compute the light spectrum using the formul{\ae} in Appendix B of \cite{Angelantonj:2024iwi} and it is reported in table \ref{tab:D5'SUSYZ2UUSpB}.
\begin{table}
\centering
\begin{tabular}{| c | c | c |}
\hline
	 { Representation} &  $\text{SO}(1,1) \times \text{SU}(2)_u \times \text{SU}(2)_v $ & { Sector} \\
	\hline
	$   \smalltableau{ \null \& \null \\}_{5'_1} +  \smalltableau{ \null \& \null \\}_{5'_2} $  & $(0, 1, 1 ) + 2 \times (\tfrac{1}{2}, 2, 1 )_L  $  & D5'-D5'
	\\
	$   \smalltableau{ \null \\ \null \\}_{5'_1} +  \smalltableau{ \null \\ \null \\}_{5'_2}  $ & $(1, 2, 2 ) + 2 \times (\tfrac{1}{2}, 1, 2 )_R$ & \ \ 
	\\
	$ ( \,  \smalltableau{ \null \\}_{5'_1} ,  \smalltableau{ \null \\}_{5'_2}) $  & $4 \times (1, 1, 1 ) + 2 \times (\frac{1}{2}, 2, 1 )_R $ &  \ \ 
	\\
	 $   ( \,  \smalltableau{ \null \\}_{5'_1} ,  \smalltableau{ \null \\}_{5'_2})  $  & $ 2 (\frac{1}{2},  1,2 )_L $ &  \ \
  \\
 \hline
	$ ( \,  \smalltableau{ \null \\}_{5'_1} ,  \smalltableau{ \null \\}_{9})   +  \, (  \, \smalltableau{ \null \\}_{5'_2} , \smalltableau{ \null \\}_{9}) $  & $ (1, 2, 1 ) + 2 \times (\frac{1}{2},  1, 1 )_R $ & D5'-D9  
 \\
	$( \,  \smalltableau{ \null \\}_{5'_1} ,  \smalltableau{ \null \\}_{9})   +  \, (  \, \smalltableau{ \null \\}_{5'_2} , \smalltableau{ \null \\}_{9}) $  & $ 2 \times (\frac{1}{2},  1, 1 )_L $ & \ \
	\\ 
 \hline
	$2^{b/2}  ( \,  \smalltableau{ \null \\}_{5'_1} ,  \smalltableau{ \null \\}_{5_{1}})   + 2^{b/2}  ( \,  \smalltableau{ \null \\}_{5'_{2}} ,  \smalltableau{ \null \\}_{5_2}) $  & $  (\frac{1}{2},  1, 1 )_L $ & D5'-D5 
 \\
 \hline
\end{tabular}
\caption{ Spectrum for probe D5' branes yielding a $\text{USp}(r_1) \times \text{USp}(r_2)$ gauge group.}
\label{tab:D5'SUSYZ2UUSpB}
\end{table}
The anomaly can be then straightforwardly computed and reads
\begin{equation}
\begin{split} \label{eq:I4D5'UxUSpSUSYZ2B}
    I_4= \tfrac{1}{2} & \left (  (r_1 + r_2) \, \text{tr} R^2 + 2^{b/2-1}  r_1 \,\text{tr} F_{ 5,1}^2 + 2^{b/2-1}  r_2 \,\text{tr} F_{5,2}^2 - (r_1-r_2)^2 \chi(N) \right ) \, ,
\end{split}
\end{equation}
allowing to cancel the bulk inflow if the charge is
\begin{equation}
    Q= \left ( \tfrac{2^{b/2}}{2\sqrt{2}}(r_1+r_2),  \tfrac{2^{b/2}}{2\sqrt{2}}(r_1+r_2); -\tfrac{2^{b/2}}{4}(r_1-r_2) \boldsymbol{1}_{2^{4-b}}; \boldsymbol{0}_{n_T-2^{4-b}-1} \right ) \, ,
\end{equation}
where we have introduced the vector $\boldsymbol{1}_d$ with components $( \boldsymbol{1}_d )_j= 1$.
Indeed, consistently with the expression reported in \eqref{eq:I4D5'UxUSpSUSYZ2B}, the levels of the algebras are given by
\begin{equation}
    k_1= 0 \, , \qquad k_2=  2^{b/2-1} r_1 \, , \qquad k_3=2^{b/2-1} r_2 \, .
\end{equation}
Notice that the algebra is always realised on the left movers, implying for the minimal choice $r_1=2 , r_2=0$
\begin{equation}
    \begin{split}
    &\sum_{i \, | \, k_i <0} \frac{|k_i| \ \text{dim} G_i }{|k_i| + h_i^{\vee}}= \frac{2^{4}(2^{4-b/2}+1)}{2 + 2^{4-b/2}+  2^{1+b/2}} \leq c_{\text{L}}-4_{\text{CM}}=16 + 96 \cdot 2^{-b/2} \, ,
    \\
     &\sum_{i \, | \, k_i \geq 0} \frac{k_i \ \text{dim} G_i }{k_i + h_i^{\vee}}= 0 \leq c_{\text{R}}-6_{\text{CM}}= 6+ 96 \cdot 2^{-b/2} \, .
\end{split}
\end{equation}

\noindent We can move on to discuss the BSB construction with gauge group
$\text{U}(4) \times \text{U}(4) \times  \text{U}(4) \times \text{U}(4)$ of Section \ref{SSec:Z4BBSB}, for which the anomaly lattice is spanned by
\begin{equation} \label{eq:bsb4coeffB}
	\begin{aligned}
		&a=\left (- \tfrac12, -\tfrac32; \boldsymbol{0}_4; \boldsymbol{0}_2 \right ) \, ,
		\\
		&b_1= \left (1, 1; \boldsymbol{\delta}^1_4; \boldsymbol{0}_2 \right ) \, ,
		\\
		&b_2= \left (1, 1; -\boldsymbol{\delta}^1_4; \boldsymbol{0}_2\right ) \, ,
		\\
		&b_3=\left (-\tfrac12,  \tfrac12; \boldsymbol{\delta}^1_4; \boldsymbol{0}_2 \right ) \, ,
		\\
		&b_4= \left (-\tfrac12, \tfrac12; -\boldsymbol{\delta}^1_4; \boldsymbol{0}_2 \right ) \, .
	\end{aligned}
\end{equation}
The D1 branes sit in $\mathbb{Z}_4$ fixed points which in the present case support O5$_-$ planes. As a result, they carry a $\text{U}(r_1)\times \text{U}(r_2)$ gauge group whose light spectrum is described in table  \ref{tab:D1BSBZ4B}.   
\begin{table}
\centering
\begin{tabular}{| c | c| c| }
\hline
	 { Representation} &  $\text{SO}(1,1) \times \text{SU}(2)_u \times \text{SU}(2)_v $ & { Sector}
  \\
	\hline
	$ ( \smalltableau{ \null \\ } \times \overline{\smalltableau{\null \\}})_{1_1}+  ( \smalltableau{ \null \\ } \times \overline{\smalltableau{\null \\}} )_{1_2}$& $(0, 1, 1 ) + 2 \times (\tfrac{1}{2}, 2, 1 )_\text{L}  $ & D1-D1 
	\\
		$ ( \smalltableau{ \null \\ } \times \overline{\smalltableau{\null \\}}  )_{1_1} +  ( \smalltableau{ \null \\ } \times \overline{\smalltableau{\null \\}}  )_{1_2}$ & $(1, 2, 2 ) + 2 \times (\tfrac{1}{2}, 1, 2 )_\text{R}$ & \ \
	\\
	$ ( \, \smalltableau{\null \\}_{1_1}, \overline{\smalltableau{\null \\}}_{1_2} ) + \smalltableau{ \null \& \null \\}_{1_1} + \overline{\smalltableau{ \null \& \null \\}}_{1_2} $ &   $4 \times (1, 1, 1 ) + 2 \times (\frac{1}{2}, 2, 1 )_\text{R} $ & \ \  
	\\
	$ ( \, \smalltableau{\null \\}_{1_1}, \overline{\smalltableau{\null \\}}_{1_2} ) +  \smalltableau{ \null \\ \null \\}_{1_1} + \overline{\smalltableau{ \null \\ \null \\}}_{1_2} $ & $2 \times (\frac{1}{2},  1,2 )_\text{L} $ & \ \ 
 \\
 \hline
	$ 2(\, \smalltableau{\null \\}_{1_1} , \overline{\smalltableau{\null \\}}_{9_1} )+ 2 ( \, \smalltableau{\null \\}_{1_2},\overline{\smalltableau{\null \\}}_{9_2}  ) $ & $2 \times (\frac{1}{2},  1, 1 )_\text{L} $ & D1-D9
\\
\hline
	$ (\, \smalltableau{\null \\}_{1_1},  \overline{\smalltableau{\null \\}}_{\bar 5_1}  )+ (\, \smalltableau{\null \\}_{1_2} , \overline{\smalltableau{\null \\}}_{\bar 5_2} ) $ & $2 \times  (1, 1, 2 ) + 4 \times (\frac{1}{2},  1, 1 )_\text{L} $ & D1-$\overline{\text{D5}}$
	\\
	$ (\, \smalltableau{\null \\}_{1_1}, \smalltableau{\null\\}_{\bar 5_1} ) + (\, \overline{\smalltableau{\null \\}}_{1_1}, \smalltableau{\null \\}_{\bar 5_2} ) $ & $ 2 \times (\frac{1}{2},  1, 1 )_\text{R} $ & \ \
 \\
 $ (\, \smalltableau{\null \\}_{1_2}, \smalltableau{\null\\}_{5_1} ) + (\, \overline{\smalltableau{\null \\}}_{1_2}, \smalltableau{\null \\}_{5_2} ) $ & $ 2 \times (\frac{1}{2},  1, 1 )_\text{R} $ & \ \
	\\
	\hline
\end{tabular}
\caption{D1 branes supporting a $\text{U}(r_1) \times \text{U}(r_2)$ gauge group placed on a $\mathbb{Z}_4$ fixed point for the BSB $T^4/\mathbb{Z}_4$ vacuum with a rank $2$ Kalb-Ramond field.}
\label{tab:D1BSBZ4B}
\end{table}
The microscopic anomaly is thus encoded into
   \begin{equation}
      \begin{aligned}
		 I_4=\tfrac{1}{2} & \left ( -(r_1+ r_2 ) \text{tr}R^2   - \chi(N) (r_1-r_2)^2 \right.
   \\
   & \left. +  2 r_1 \, \text{tr} F_{9,1}^2 +  2 r_2 \, \text{tr} F_{9,2}^2  + (r_1-r_2) \ \text{tr} F_{5,1}^2 + (r_2-r_1) \ \text{tr} F_{5,2}^2 \right ) \, ,
   \end{aligned}
   \end{equation}
which is cancelled by the bulk inflow by choosing the charge vector 
   \begin{equation}
       Q=\left ( \tfrac{r_1+r_2}{2},- \tfrac{r_1+r_2}{2};(r_1-r_2)\boldsymbol{\delta}^1_4 ; \boldsymbol{0}_2 \right ) \, .
   \end{equation}
Furthermore, from the anomaly polynomial, we can read the level of the Ka$\check{\text{c}}$-Moody algebras realised on the (anti)holomorphic sector, yielding
\begin{equation}
    k_1= 2 r_1 \, , \qquad k_2=2 r_2 \, , \qquad k_3=r_1-r_2 \, , \qquad r_4=r_2-r_1 \, .
\end{equation}
Notice the presence of the $2$ for the level corresponding to D9 algebras which correspond to the extra $2^{b/2}$ from the presence of the $B$-field.
Choosing for instance the minimal configuration $r_1=1$ and $r_2=0$, we can evaluate the unitarity constraints which yields
\begin{equation} \label{eq:SUSYZ2U}
\begin{split}
   & \sum_{i \, | \, k_i \geq 0} \frac{k_i \ \text{dim} G_i }{k_i + h_i^{\vee}}= 6 + 3  < c_{\text{L}}-4_{\text{CM}}=36 \, , 
    \\
   & \sum_{i \, | \, k_i <0} \frac{|k_i| \ \text{dim} G_i }{|k_i| + h_i^{\vee}}=3 < c_{\text{R}}-6_{\text{CM}}=30 \, . 
\end{split}
\end{equation}

\noindent A similar discussion applies for the case of D5' branes, yielding an $\text{USp}(r_1) \times \text{Usp}(r_2)$ gauge group. The light spectrum can be straightforwardly computed and is reported in table \ref{tab:D5'BSBZ4B}. 
\begin{table}
\centering
\begin{tabular}{| c | c| c| }
\hline
	 { Representation} &  $\text{SO}(1,1) \times \text{SU}(2)_u \times \text{SU}(2)_v $ & { Sector}
  \\
	\hline
 $ ( \smalltableau{ \null \\ } \times \overline{\smalltableau{\null \\}} )_{5'_1}+  ( \smalltableau{ \null \\ } \times \overline{\smalltableau{\null \\}} )_{5'_2}$& $(0, 1, 1 ) + 2 \times (\tfrac{1}{2}, 2, 1 )_\text{L}  $ & D1-D1 
	\\
		$ ( \smalltableau{ \null \\ } \times \overline{\smalltableau{\null \\}} )_{5'_1}+  (   \smalltableau{ \null \\ } \times \overline{\smalltableau{\null \\}}  )_{5'_2}$ & $(1, 2, 2 ) + 2 \times (\tfrac{1}{2}, 1, 2 )_\text{R}$ & \ \
	\\
	$ (\, \smalltableau{\null \\}_{5'_1}, \overline{\smalltableau{\null \\}}_{5'_2} ) +  \smalltableau{ \null \& \null \\}_{5'_1} + \overline{\smalltableau{ \null \& \null \\}}_{5'_2} $ &   $4 \times (1, 1, 1 ) + 2 \times (\frac{1}{2}, 2, 1 )_\text{R} $ & \ \  
	\\
	$ (\, \smalltableau{\null \\}_{5'_1}, \overline{\smalltableau{\null \\}}_{5'_2} ) +  \smalltableau{ \null \\ \null \\}_{5'_1} + \overline{\smalltableau{ \null \\ \null \\}}_{5'_2} $ & $2 \times (\frac{1}{2},  1,2 )_\text{L} $ & \ \ 
 \\
\hline
	$ (\, \smalltableau{\null \\}_{5'_1},  \overline{\smalltableau{\null \\}}_{9_1} )+ (\, \smalltableau{\null \\}_{5'_2} ,\overline{\smalltableau{\null \\}}_{9_2} ) $ & $2 \times (1, 2, 1 ) + 4 \times (\frac{1}{2},  1, 1 )_\text{R} $ & D5$'$-D9
	\\
	$ (\, \smalltableau{\null \\}_{5'_1},  \smalltableau{\null\\}_{9_1}  ) + (\, \overline{\smalltableau{\null \\}}_{5'_1}, \smalltableau{\null \\}_{9_2} )$ & $ 2 \times (\frac{1}{2},  1, 1 )_\text{L} $ & \ \
 \\
 $  (\, \smalltableau{\null \\}_{5'_2}, \smalltableau{\null\\}_{9_1} ) + (\,\overline{\smalltableau{\null \\}}_{5'_2}, \smalltableau{\null \\}_{9_2}  ) $ & $ 2 \times (\frac{1}{2},  1, 1 )_\text{L} $ & \ \
\\
\hline
	$   2 (\, \smalltableau{\null \\}_{5'_1},\overline{\smalltableau{\null \\}}_{\bar 5_{1}}  )+ 2 (\, \smalltableau{\null \\}_{5'_2} , \overline{\smalltableau{\null \\}}_{\bar 5_{2}} ) $ & $2 \times (\frac{1}{2},  1, 1 )_\text{R} $ & D5$'$-$\overline{\text{D5}}$
	\\
	\hline
\end{tabular}
\caption{D5$'$ branes supporting a $\text{U}(r_1) \times \text{U}(r_2)$ gauge group for the BSB $T^4/\mathbb{Z}_4$ vacuum with a rank $2$ Kalb-Ramond field.}
\label{tab:D5'BSBZ4B}
\end{table}
In this scenario, the anomaly polynomial reads
   \begin{equation}
      \begin{aligned}
        I_4=\tfrac{1}{2} &\left ( -(r_1 + r_2) \text{tr}R^2  - \chi(N) \left (r_1- r_2 \right )^2  \right.
		\\
		&\left. + (r_2-r_1) \, \text{tr} F_{9,1}^2 + (r_1-r_2) \, \text{tr} F_{9,2}^2 - 2  r_1 \, \text{tr} F_{5,1}^2 - 2 r_2 \, \text{tr} F_{5,2}^2 \right ) \, ,
  \end{aligned}
   \end{equation}
with charge vector cancelling the bulk inflow given by 
   \begin{equation}
      Q = \left ( \tfrac{r_1+r_2}{2}, \tfrac{r_1+r_2}{2}; \tfrac{r_2-r_1}{2} \boldsymbol{1}_4 ; \boldsymbol{0}_2 \right ) \, .
   \end{equation}
The anomaly polynomial allows thus to compute the level of the Ka$\check{\text{c}}$-Moody algebras arising in the IR and are given by 
\begin{equation}
    k_1= r_2-r_1 \, , \qquad k_2= r_1-r_2 \, , \qquad k_3=- 2 r_1 \, , \qquad k_4=-2 r_2
\end{equation}
We can evaluate the constraints dictated by unitarity by choosing for instance $r_1=2$ and $ r_2=0$, which yields
\begin{equation}
    \begin{split}
    &\sum_{i \, | \, k_i <0} \frac{|k_i| \ \text{dim} G_i }{|k_i| + h_i^{\vee}}= 5 < c_{\text{L}}-4_{\text{CM}}=28 \, ,
    \\
     &\sum_{i \, | \, k_i \geq 0} \frac{k_i \ \text{dim} G_i }{k_i + h_i^{\vee}}= \frac{25}{2} < c_{\text{R}}-6_{\text{CM}}= 38\, .
\end{split}
\end{equation}

\section{Conclusions} \label{Sec:Conclusions}

In the present paper, we have discussed the possibility of introducing a non-vanishing background $B$-field for BSB and supersymmetric orientifolds built upon the $T^4/\mathbb{Z}_N$ orbifolds. We have completed the classification of all the possible solutions for the $T^4/\mathbb{Z}_2$ orbifold initiated in \cite{Angelantonj:1999jh} along the lines of \cite{Angelantonj:2024iwi}. A new family of solutions shows up, which is compatible with R-R tadpole conditions for real Chan-Paton factors arising from placing D9 and D5 or $\overline{{\text{D5}}}$ branes on the top of O5$_+$ planes. In addition, we have shown how the presence of a non-trivial background field also supports a hybrid situation in which one of the Chan-Paton factors is complex, while the other one is real. The anomaly polynomials of these models have been computed and furthermore, the unitarity conditions for the theory living on the world-volume of string defects allowed by anomaly inflow have been verified. The possibility of introducing a rank $b$ Kalb-Ramond field has been then extended to the other available $T^4/\mathbb{Z}_N$ orbifolds. While for the $T^4/\mathbb{Z}_6$ orbifold and the $T^4/\mathbb{Z}_4$ orbifold with a rank $4$ Kalb-Ramond field the number of D-branes to be introduced would be fractional, explicit solutions have been found for the supersymmetric and BSB $T^4/\mathbb{Z}_4$ orientifolds with a rank $2$ Kalb-Ramond field. 
However, a cancellation of R-R tadpoles cannot be achieved for an arbitrary configuration of orientifold planes. Indeed, a correct interpretation of the tree-level amplitudes requires orientifold planes placed over all the $\mathbb{Z}_4$ fixed points to carry a vanishing twisted charge. When supersymmetry is linearly realised, the analysis completes the discussion first initiated in \cite{Kakushadze:1998bw}, while for BSB vacua, introducing a rank $2$ Kalb-Ramond field relaxes the rigidity of the model. Indeed, O5$_+$ planes on $\mathbb{Z}_4$ fixed points carry non-trivial twisted charges and thus forbid the possibility to recombine $\overline{\text{D5}}$ branes. The non-trivial background value for the $B$-field replaces O5$_+$ planes with O5$_-$ ones that are uncharged with respect to the twisted six-forms localised on the $\mathbb{Z}_4$ fixed points. As a result, recombination of branes is now allowed, allowing to deform the model to other vacua. 
These new models have been further checked to be consistent by verifying the cancellation of local anomalies and the unitarity constraints arising from the introduction of string defects.      

\noindent  To iterate, the introduction of the $B$-field seems to imply strong constraints on the structure of the vacuum. Indeed, when the orientifold amplitudes are built from higher-order point group orbifolds, the cancellation of local anomalies, as well as a consistent interpretation of the vacuum-to-vacuum and tree-level amplitudes, forbids the presence of a non-vanishing rank $b$ Kalb-Ramond field for the $T^4/\mathbb{Z}_6$  case or for a rank $4$ $B$-field for the $T^4/\mathbb{Z}_4$ case. In addition, the configuration of  orientifold planes for the allowed cases is not arbitrary, but follows a specific setting. Therefore, in such a setting, it would be interesting to understand if these features underlie dynamical or topological obstructions along the lines of \cite{Witten:1997bs, Sen:1997pm, Bachas:2008jv}. In the latter case, this would mean understanding if the action of the orientifold projection combined with the orbifold point group constrains the characteristic class in equivariant $\mathbb{Z}_2$ cohomology to be trivial for the $T^4/\mathbb{Z}_6$ case and to support at most a rank $2$ $B$-field for the $T^4/\mathbb{Z}_4$ case. This would lead to a clear understanding of the features of these vacua.   

\noindent Furthermore, a more complete analysis of the consistency of these models would be provided by verifying the cancellation of Dai-Freed anomalies \cite{Dai:1994kq, Garcia-Etxebarria:2018ajm, Debray:2021vob, Debray:2023yrs} along the lines of \cite{Basile:2023knk, Basile:2023zng}.

\section*{Acknowledgements}

The author would like to thank Carlo Angelantonj, Cezar Condeescu and Emilian Dudas for discussions and collaboration on related projects. Furthermore, the author would like to thank Carlo Angelantonj for useful clarifications and comments on vacua with a non-vanishing $B$-field and Ivano Basile for interesting and useful discussions about the topological aspects of the analysis. The author is grateful to the IPHT of Paris Saclay and CERN for their hospitality while this project was underway.

\appendix

\section{Useful formul{\ae} for the computation of the partition functions}
\label{App:ZN}

Throughout all the paper we have described the partition functions for the $T^4/\mathbb{Z}_N$ orientifolds by using the building blocks encoding the orbifold action. This Appendix is devoted to give the explicit expressions for such terms in order to make the exposition self-contained, gathering the essential information contained in the Appendices of \cite{Angelantonj:2024iwi}. We refer to the latter for a more detailed presentation. 

\noindent The modular blocks employed in our discussion can be written as
\begin{equation} \label{eq:modblock}
\begin{aligned}
    T_{s} \big[ {\textstyle{\alpha \atop \beta}}\big]  &= \text{tr}_\alpha \, \left( g^\beta\, P^s_\text{GSO}\, q^{L_0-c/24} \right)
    \\
    &= \sum_{a,b=0,\frac{1}{2}} \tfrac{1}{2}\eta^s_{a,b} \, 
\prod_{i=1}^4 \frac{\theta \big[{\textstyle{a+\alpha_i \atop b+\beta_i}}\big]}{\theta \big[{\textstyle{1/2+\alpha_i \atop 1/2+\beta_i}}\big]}\, d_{\alpha_i , \beta_i} \,,
\end{aligned}
\end{equation}
where $\eta$ and $\theta \big[{\textstyle{a \atop b}}\big]$ indicate the Dedekind eta and Jacobi theta function with upper, $a$, and lower, $b$, characteristics, respectively. We also adopt the convention for which $q=e^{2\pi i \tau}$. The vectors $\boldsymbol{\alpha} = (0,0,\alpha/N , -\alpha/N)$ and $\boldsymbol{\beta} = (0,0,\beta/N , -\beta/N)$ are associated to the orbifold action and label the $\alpha$ twisted sector with the insertion of the group element $g^\beta$ in the trace. The index $s$ reflects the phase $\eta^s_{a,b}$ which defines the GSO projection. 
Finally, the modular blocks depend on the coefficients given by $d_{\mu , \nu } = 2\, \sin  (\pi \nu )$ if $\mu =0$ or $1$ if $\mu \neq 0$, which removes the numerical degeneracy from the theta functions associated to the compact bosons.

\noindent The spectrum can be easily extracted by diagonalising the modular blocks as
\begin{equation}
T_s \big[ {\textstyle{\alpha \atop \beta}}\big] = \sum_{\gamma =0}^{N-1} e^{2 i \pi \beta \gamma/N} \, \tau^s_{\alpha , \gamma} 
\end{equation}
where we have defined the combinations of characters $\tau_{\alpha\, \beta}^s$ that are eigenstates under the orbifold action. Different GSO projections lead to different characters. 

\noindent When the type IIB GSO projection is employed, the phase $\eta_{a,b}^s$ corresponds to
\begin{equation}
\eta^B_{a,b} = (-1)^{2a+2b+4ab}\,, 
\end{equation}
inducing the characters to form representations of the ${\mathcal N}=(1,0)$ supersymmetry algebra. We can extract the massless contributions which appear only in \cite{Angelantonj:2024iwi} in terms of the little group $\text{SO}(4)$ characters \cite{Angelantonj:2002ct}
\begin{equation}
\tau^B_{0,0} \sim V_4 - 2 S_4\,, \qquad \tau^B_{0,1}\,,\ \tau^B_{0,N-1}\,,\ \tau^B_{\alpha ,0} \sim 2O_4 - C_4 \qquad (\alpha\neq 0)\,,
\end{equation}
while all the other characters contribute only to the massive terms. As described in \cite{Angelantonj:2024iwi}, the $N=2$ case identifies $\tau^B_{0,1} $ with $ \tau^B_{0,N-1}$, so that its content is doubled.

\noindent However, when dealing with the vacuum-to-vacuum amplitudes describing strings stretched between D9 and $\overline{\text{D5}}$ branes a different GSO projection has to be used. This corresponds to 
\begin{equation}
\eta^{\hat B}_{a,b} = (-1)^{2a +4ab}\,,
\end{equation}
resulting in orbifold characters containing states which do not form super multiplets anymore. The massless contribution to the spectrum is given by
\begin{equation} \label{eq:masslesschar}
    \tau^{\hat B}_{N/2,0} \sim - S_4 \, , \qquad \tau^{\hat B}_{N/2,1} \, ,\tau^{\hat B}_{N/2,N-1} \sim  O_4
\end{equation}
while the other untwisted and twisted characters do not enter in orientifold amplitudes and thus have no use for us. Again for the $N=2$, the $\tau^{\hat B}_{N/2,1}$ character and the $  \tau^{\hat B}_{N/2,N-1}$ one are joined together to form a couple of real scalars.

\noindent These terms are enough to describe the partition functions entering the vacuum-to-vacuum amplitudes. For what concerns the tree-level amplitudes, the relevant modular blocks arise from the modular properties of the Dedekind eta and Jacobi theta functions. The relevant modular blocks appearing in Section \ref{Sec:6dorientifold} are given by 
\begin{equation} 
\begin{aligned}\label{eq:modB}
    B_B \big[ {\textstyle{\alpha \atop \beta}}\big]  = \sum_{a=0 \atop b=0,\frac{1}{2}} \tfrac{1}{2}\eta^B_{a,b} \, 
\prod_{i=1}^4 \frac{\theta \big[{\textstyle{a+\alpha_i \atop b+\beta_i}}\big]}{\theta \big[{\textstyle{1/2+\alpha_i \atop 1/2+\beta_i}}\big]}\, d_{\alpha_i , \beta_i} \,,
\end{aligned}
\end{equation}
and 
\begin{equation} 
\begin{aligned}\label{eq:modF}
    F_B \big[ {\textstyle{\alpha \atop \beta}}\big]  = - \sum_{a=\frac{1}{2} \atop b=0,\frac{1}{2}} \tfrac{1}{2}\eta^B_{a,b} \, 
\prod_{i=1}^4 \frac{\theta \big[{\textstyle{a+\alpha_i \atop b+\beta_i}}\big]}{\theta \big[{\textstyle{1/2+\alpha_i \atop 1/2+\beta_i}}\big]}\, d_{\alpha_i , \beta_i} \,,
\end{aligned}
\end{equation}
which encode the bosonic and fermionic pieces of \eqref{eq:modblock}. We can diagonalise \eqref{eq:modB} and \eqref{eq:modF}, as we did for the  \eqref{eq:modblock}, by 
\begin{equation}
B_B \big[ {\textstyle{\alpha \atop \beta}}\big] = \sum_{\gamma =0}^{N-1} e^{2 i \pi \beta \gamma/N} \, \beta^B_{\alpha , \gamma} \, , \qquad 
F_B \big[ {\textstyle{\alpha \atop \beta}}\big] = \sum_{\gamma =0}^{N-1} e^{2 i \pi \beta \gamma/N} \, \varphi^B_{\alpha , \gamma} \, ,
\end{equation}
where the state content can be extracted from \eqref{eq:masslesschar} by noticing that $\tau^B_{\alpha,\gamma}= \beta^B_{\alpha, \gamma} - \varphi^B_{\alpha, \gamma}$.

\newpage 

\printbibliography


\end{document}